\documentclass[preprint,12pt]{aastex}  
\usepackage{epsfig}

\def\cm{\ifmmode {\rm cm}^{-1} \else cm$^{-1}$ \fi}
\def\s{\ifmmode {\rm s}^{-1} \else s$^{-1}$ \fi}
\def\cc{\ifmmode {\rm cm}^{-3} \else cm$^{-3}$ \fi}
\def\cs{\ifmmode {\rm cm}^{-2} \else cm$^{-2}$ \fi}
\def\g{\ifmmode \gamma \else $\gamma$\fi}
\def\G{\ifmmode \Gamma \else $\Gamma$\fi}
\def\Gs{\ifmmode \Gamma~ \else $\Gamma~$\fi}

\def\gc{\ifmmode \gamma_{\rm c} \else $\gamma_{\rm c}$ \fi}
\def\sw{Schwarzschild~}
\def\gsim{\mathrel{\raise.5ex\hbox{$>$}\mkern-14mu
             \lower0.6ex\hbox{$\sim$}}}
\def\lsim{\mathrel{\raise.3ex\hbox{$<$}\mkern-14mu
             \lower0.6ex\hbox{$\sim$}}}
\def\simless{\mathbin{\lower 3pt\hbox
     {$\rlap{\raise 5pt\hbox{$\char'074$}}\mathchar"7218$}}}   
\def\simmore{\mathbin{\lower 3pt\hbox
     {$\rlap{\raise 5pt\hbox{$\char'076$}}\mathchar"7218$}}}   
\def\Msun{M_\odot}                                
\def\4u{4U 1728--34}
\def\deg{^\circ}

\lefthead{et al.} \righthead{\4u}

\shorttitle{QPOs from Random X-ray Bursts}

\shortauthors{Fukumura, Kazanas, \& Stephenson}

\begin{document}

\title{QPOs from Random X-ray Bursts around Rotating Black Holes}


%
\author{\textsc{Keigo Fukumura}\altaffilmark{1,2,3},
\textsc{Demosthenes Kazanas}\altaffilmark{3} \textsc{and}
\textsc{Gordon Stephenson}\altaffilmark{4} }

\altaffiltext{1}{Email: Keigo.Fukumura@nasa.gov}

\altaffiltext{2}{University of Maryland, Baltimore County
(UMBC/CRESST), Baltimore, MD 21250}

\altaffiltext{3}{Astrophysics Science Division, NASA/Goddard Space
Flight Center, Greenbelt, MD 20771}

\altaffiltext{4}{University of California, San Diego, Mail Code
0230, La Jolla, CA 92093 }

\begin{abstract}

\baselineskip=15pt

We continue our earlier studies of quasi-periodic oscillations
(QPOs) in the power spectra of accreting, rapidly-rotating black
holes that originate from the geometric ``light echoes" of X-ray
flares occurring within the black hole ergosphere. Our present work
extends our previous treatment to three-dimensional photon emission
and orbits to allow for arbitrary latitudes in the positions of the
distant observers and the X-ray sources in place of the mainly
equatorial positions and photon orbits of the earlier consideration.
Following the trajectories of a large number of photons we calculate
the response functions of a given geometry and use them to produce
model light curves which we subsequently analyze to compute their
power spectra and autocorrelation functions. In the case of an
optically-thin environment, relevant to advection-dominated
accretion flows, we consistently find QPOs at frequencies of order
of $\sim$ kHz for stellar-mass black hole candidates while order of
$\sim$ mHz for typical active galactic nuclei ($\sim 10^7 \Msun$)
for a wide range of viewing angles ($30^{\deg}$ to $80^{\deg}$) from
X-ray sources predominantly concentrated toward the equator within
the ergosphere. As in our previous treatment, here too, the QPO
signal is produced by the frame-dragging of the photons by the
rapidly-rotating black hole, which results in photon ``bunches"
separated by constant time-lags, the result of multiple photon
orbits around the hole. Our model predicts for various
source/observer configurations the robust presence of a new class of
QPOs, which is inevitably generic to curved spacetime structure in
rotating black hole systems.


\end{abstract}


\baselineskip=15pt

\section{Introduction}

One of the more interesting findings of the X-ray timing analyses of
the light curves of accreting compact objects in the past two
decades has been the discovery of quasi-periodic oscillations
(QPOs). These are broad features in the power spectra of these
sources that range up to kHz frequencies in neutron star low-mass
X-ray binaries (see, e.g., \citealt{Klis00}) and up to tens and even
hundreds of Hz in accreting black hole binary systems (see, e.g.,
\citealt{Cui98} and \citealt{Strohmayer01a,Strohmayer01b}),
including QPO pairs with 2:3 frequency commensurability in the power
spectra of certain black hole candidates (e.g. XTE~J1550-564 and
GRO~J1655-40). {Besides galactic binary sources, the presence of
QPOs has also been reported in at least one  ultra-luminous X-ray
sources (ULX), namely NGC~5408~X-1 \citep[see][]{Strohmayer07} at
$\nu \sim 20$ mHz; in the context of active galactic nuclei (AGNs)
\citet{Gierlinski08} recently discovered a $\sim 1$ hour X-ray
periodicity in the narrow-line Seyfert 1 (NLS1) RE~J1034+396, in
which the periodicity is apparent in its light curve in distinction
with those of the galactic binary black hole systems mentioned above
that do not exhibit any such behavior.}

The nature of QPO phenomenon in accreting black hole systems is at
present largely unknown; it is also possible that this phenomenon is
not the manifestation of a unique process, but that the observed
diversity of QPO frequencies and sites may be due to a multitude of
processes specific to the particular frequency and site. For the
case of QPOs associated with black hole systems, the absence of an
underlying, rotating solid surface object, implies that plausible
explanations involve by necessity processes associated with the
surrounding accretion disks. These include, among others, accretion
disk precession due to the frame-dragging effects of a
rapidly-rotating black hole
\citep[e.g.][]{Cui98,Merloni99,Schnittman06} (also see, e.g.,
\citealt{SV98}, for the same effect invoked to account for the kHz
QPOs of Low Mass X-ray Binaries), accretion disk oscillatory modes
\citep[e.g.][for
diskoseismology]{Nowak97,Abramowicz01,Kato01,Donmez07}, or the
distribution of X-ray emitting ``blobs" spanning a limited range
around some specific radius of the Keplerian disk surrounding the
black hole \citep[e.g.][]{Karas99}. There have also been studies of
the correlations of the QPO frequencies with the properties of the
associated X-ray spectra \citep[e.g.][]{Titarchuk04}.

An altogether different notion to QPO origin has been put forward
recently by \citet[][hereafter FK08]{FK08}. These authors proposed
that high frequency QPO (HFQPO) could be produced as a consequence
of {\it light echoes} of X-ray flares occurring within the
ergosphere of a Kerr black hole; they showed that due to
frame-dragging of individual flare photons, a finite number of them
reach far away observers after an integer number of additional
orbits around the black hole to produce a geometry-induced ``light
echo" of the original flare \citep[see, also,][for a similar
discussion]{Meyer06,Bursa07}. Studying two-dimensional (2D)
near-equatorial photon propagation in a fully general relativistic
calculation, FK08 showed that dragging of inertial frames leads
approximately $15\%$ of the photons to a specific direction at
infinity (an observer) after a time lag of $\Delta t \simeq 14 \, M$
(where $M$ is black hole mass), independent of the relative position
between the observer and the photon source. The independence of the
lag on the relative phase between the observer and the source then
guarantees a second peak in the Autocorrelation Function (ACF) at
lag $\tau \simeq 14 M$, which according to Fourier analysis, results
in a prominent HFQPO signal at a frequency $\nu_{\rm QPO} \simeq
1/\Delta t \simeq 1.4 \, (10\Msun/M)$ kHz (where $\Msun$ is the
solar mass), even for a light curve that consists of flares randomly
distributed within the ergosphere. Since the closest an accretion
disk can approach the black hole is the radius of Innermost Stable
Circular Orbit (ISCO), this effect is possible only when the ISCO
lies within the black hole ergosphere, a condition that limits this
effect to black holes with dimensionless spin parameter $a/M \gsim
0.94$.

The study of FK08 showed the prominent role of frame-dragging in
producing lags independent of the relative source - observer
position; as shown there, the absence of frame dragging for flares
in a \sw geometry leads to lags which depend on the relative
observer - source position and, for random flare positions, to
absence of QPO features in the corresponding power spectra.

The original study of FK08 outlined the fundamental aspects behind
this work, namely the importance of frame-dragging and the
persistence of QPO presence, even for random positions of the X-ray
flares within the ergosphere. However, it was restricted in its
scope in that it was constrained to observers and sources near or on
the black hole equatorial plane, i.e. $\theta_s \simeq \theta_o
\simeq \pi/2$. Realistic constraints demand an enlarged study to
include, at a minimum, observers at larger latitude positions, not
least because it is expected that the column density of sources for
lines of sight near the equator are sufficiently high to preclude
propagation of photons without interaction with the surrounding
matter; this could then lead either to their absorption or to their
scattering and introduction of additional, random lags to their
trajectories erasing the otherwise potential QPO signal. The goal of
the present work is to remedy this deficiency by expanding the study
of the corresponding lags to three-dimensional (3D) geometries
allowing thus for arbitrary latitudes of both the observers as well
as of the source of photons.

We would like to stress that the present models of
``geometry-induced" QPOs (and also those of FK08) are not meant as
an account of the so-called HFQPOs presented in the literature to
date (e.g. \citealt{Cui98,Strohmayer01a,Strohmayer01b}). For one
thing, the predicted QPO frequencies are $5-10$ times higher than
the observed frequencies mentioned above (for the same black hole
mass), and, for another, they do not produce their 2:3 frequency
commensurability. Instead, the predicted QPOs should be viewed as an
altogether new feature indicating the intrusion of an accreting gas
within a black hole ergosphere, thereby providing an unequivocal
evidence for the presence of rapidly rotating black holes.

Our paper is organized as follows: In \S 2 we provide a general
description of our 3D QPO model, the details of the photon
kinematics and response function of the system, and a prescription
for constructing stochastic model light curves. In \S 3 we compute
the autocorrelation functions (ACFs) and power spectral densities
(PSDs) of the model light curves to show that they generally exhibit
the QPO features as anticipated. Finally, in \S 4 we review our
results, make contact with observations along with the limitations
of our model, and conclude by discussing prospects of future work.
{Finally, in the Appendices we discuss the details the notion of
locally isotropic emission in Kerr geometry and provide explicit
functional forms relating the photon impact parameters  to their
local emission angles.

\section{Model Details}

The present treatment follows closely that of FK08, but generalizing
the photon trajectories to 3D, thus allowing for positions of both
the source and the observer at latitudes other than near equatorial.
So, we consider photon emission in Kerr geometry of dimensionless
angular momentum of a black hole $a$, described in Boyer-Lindquist
coordinates $(r,\theta,\phi)$. As in FK08, here too we employ the
usual geometrized units ($G=c=1$ where $G$ is the gravitational
constant and $c$ is the speed of light) and normalize distance and
time with the black hole mass $M$. For example, $1M$ corresponds to
$1.5 \times 10^5 (M/\Msun)$ cm in distance and $5 \times 10^{-6}
(M/\Msun)$ sec in time. We consider the observer position at an
arbitrary location ($r_o,\theta_o,\phi_o$), while correspondingly
denoting the source coordinates as ($r_s,\theta_s,\phi_s$). This
arrangement, therefore, allows for source locations not only at
random azimuths but also at random polar angles. In thin accretion
disks, the flares are likely confined to small latitudes, i.e.
$\theta_s \simeq \pi/2$, but for quasi-spherical accreting flows
such as advection-dominated accretion flows \citep[ADAFs;
e.g.][]{NY94}, the overall source latitude can be much higher. As
discussed in following subsections, in this last case, one can, in
addition, weigh the emission at each latitude to fit one's preferred
flow model.

\subsection{X-ray Flares}

\begin{figure}[t]
\epsscale{1} \plottwo{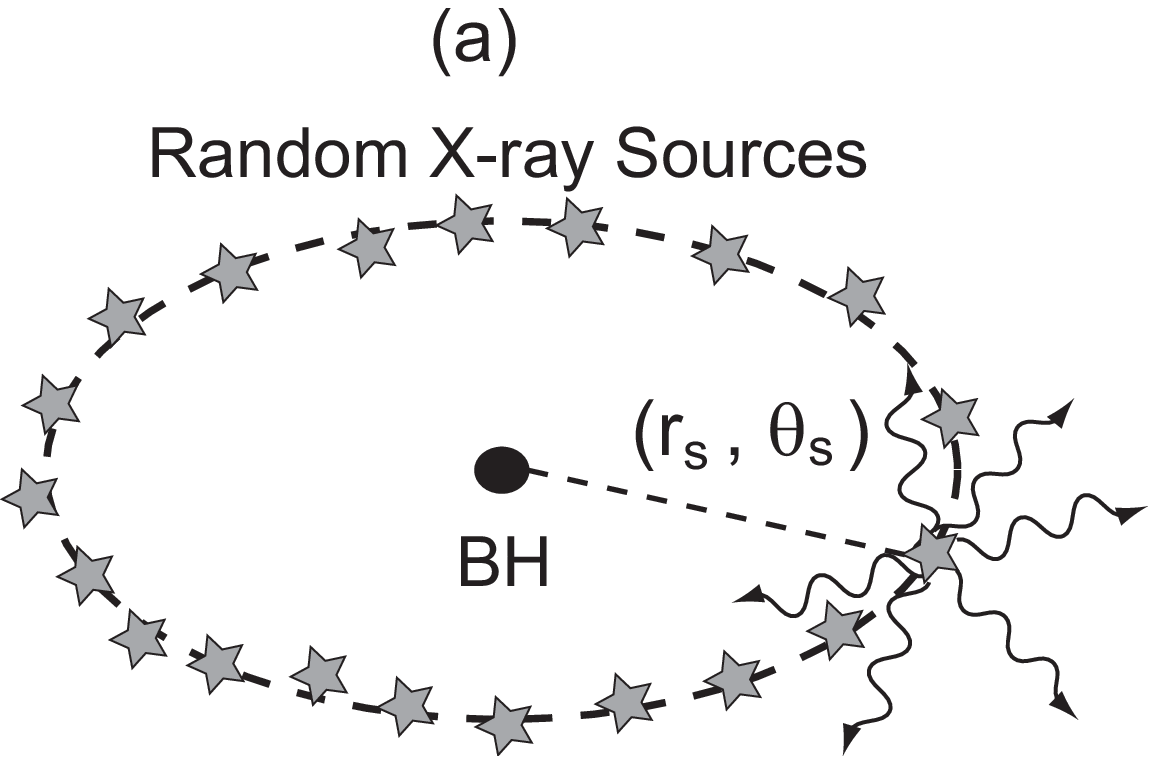}{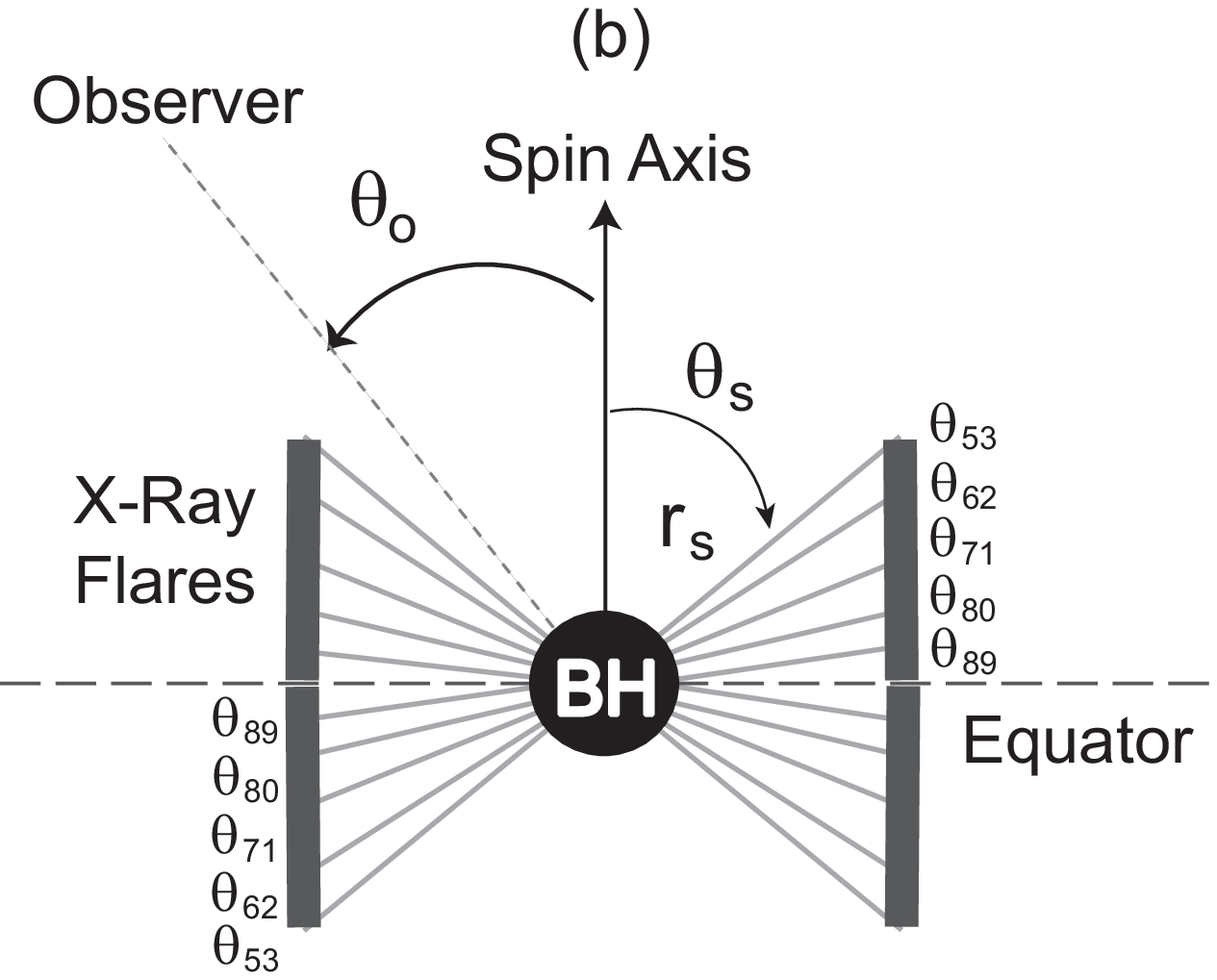} \caption{(a) Schematic view
of the system considered in this paper. (b) Poloidal projection of
the geometrical setup in this problem. \label{fig:geometry}}
\end{figure}

The X-ray emission of the sources we consider is regarded as the
incoherent sum of a large number of instantaneous (i.e. short
compared to the dynamical time) flares. We assume the typical
duration of such a flare to be $\sim 1M$, presumably originating
from local magnetic reconnections \citep[e.g.][]{SZ94} or standing
shocks in relativistic magnetized accretion
\citep[e.g.][]{Takahashi02,Fukumura07}. Concerning generation of
X-ray flares within the ergosphere, \citet{Karas08} have attributed
it to magnetic reconnection induced by frame-dragging, while
\citet{Koide08} have considered magnetic reconnection within the
ergosphere as a means of extracting energy from the black hole
rotation. In general, these X-ray flares do not necessarily have to
be confined exactly to the equatorial plane, and for this reason, in
this work, we relax this restriction on the source vertical
location. 
Therefore, in the specific cases that the emitting plasma is
optically thin, the observed signal will contain contributions from
sources in both {\it upper} and {\it lower} hemispheres.

It is important to mention that in this model both polar and
azimuthal positions ($\theta_s,\phi_s$) of each X-ray flash are
randomly assumed (as long as they lie within or close to the
ergosphere) while their cylindrical radius $r_s \sin \theta_s$ is
fixed for simplicity to the radius of the ISCO of a Kerr hole with
$a/M \simeq 0.99$, i.e. to $r_s \sin \theta_s \simeq 1.455M$. One of
our primary interests is thus to probe the dependence on the source
{and the observer} latitudes as illustrated in
Figure~\ref{fig:geometry}b. To avoid the introduction of artificial
coherent signals associated with orbital motion of these emitting
sources, we also consider the flares to be randomly distributed in
time in the following fashion: each X-ray burst occurs with a
Poisson distribution of mean value equal to the Keplerian orbital
timescale; $\Delta T \equiv T_{\rm orb} |\ln \{\textrm{rnd}(0,1)\}|$
where $T_{\rm orb}$ is the orbital time of the source at
$(r_s,\theta_s)$ and $\textrm{rnd(0,1)}$ is a random number between
0 and 1. We define that $T_{\rm orb} \equiv 2\pi \{(r_s \sin
\theta_s)^{3/2}+a M^{1/2}\}/M^{1/2}$.

The orbiting X-ray emitting plasma is assumed to be in Keplerian
motion, either below or above an equatorial plane at $\theta_{s,i}$
where $i$ denotes the i-th source, i.e. rotating with the local
Keplerian angular velocity (modified to take into account the black
hole spin $a$)
\begin{eqnarray}
\Omega_s = \left[ \frac{M^{1/2}} {(r \sin \theta)^{3/2} + a M^{1/2}}
\right]_{\rm source} \ , \label{eq:Omega}
\end{eqnarray}
with all the quantities being evaluated at the source position. In
the lab frame or the locally non-rotating reference frame (LNRF)
each source has a total three-velocity of
\begin{eqnarray}
v_s &=& \left[\frac{A \sin \theta}{\Sigma \Delta^{1/2}}
(\Omega-\omega) \right]_{\rm source} \ , \label{eq:vphi}
\end{eqnarray}
where we define $\Delta \equiv r^2-2Mr+a^2$, $\Sigma \equiv r^2+a^2
\cos^2 \theta$, $A \equiv (r^2+a^2)^2-a^2 \Delta \sin^2 \theta$, and
frame-dragging of the local inertial frame is denoted by $\omega
\equiv 2 Mra/A$. In the LNRF, as one can see, rotation of a local
inertial frame $\omega$ has been subtracted by definition.

\subsection{Null Geodesics}

Neglecting external interactions between photons and particles, each
null geodesic (i.e. photon orbit) is uniquely characterized by two
constants of motion, $\xi$ and $\eta$, where $\xi$ is the axial
component of angular momentum and $\eta$ closely related to its
polar component
\citep[e.g.][]{Bardeen72,Chandra83}. For given $\eta$ and $\xi$, a
photon orbit in Kerr geometry is generally governed by the following
equations of motion \citep[e.g.][]{Bardeen72,Chandra83};
\begin{eqnarray}
\dot{t} &=& \frac{(r^2+a^2)\Xi +a \Delta \xi-a^2 \Delta \sin^2
\theta}{\Delta \Sigma} \ , \label{eq:t}
\\
\dot{r}^2 \Sigma^2 &=& \Xi^2 - \Delta \{\eta+(\xi-a)^2\} \ ,
\label{eq:r}
\\
\dot{\theta} \Sigma^2 &=& \eta+(a^2-\xi^2 \csc^2 \theta) \cos^2
\theta \ , \label{eq:theta}
\\
\dot{\phi} &=& \frac{a(2Mr-a\xi)+\xi \Delta \csc^2 \theta}{\Delta
\Sigma} \ , \label{eq:phi}
\end{eqnarray}
where $\Xi \equiv r^2+a^2-a\xi$ and an overdot denotes derivative
with respect to the affine parameter. Note that equations
(\ref{eq:r}) and (\ref{eq:theta}) provide only the squares of
$\dot{r}$ and $\dot{\theta}$. In order to avoid the issue of
determination of appropriate signs for $\dot{r}$ and $\dot{\theta}$,
for the actual numerical determination of the orbits we use instead
the second derivatives of equations~(\ref{eq:r}) and
(\ref{eq:theta})
\begin{eqnarray}
\ddot{r} &=& \frac{2 r \dot{r} \Xi  - \dot{r} \left(r-M \right)
\left\{\eta+(\xi-a)^2\right\}- \Sigma \left\{2r \dot{r}-a^2
\sin(2\theta) \dot{\theta} \right\} \dot{r}^2}{\dot{r} \Sigma^2} \ ,
\label{eq:r2}
\\
\ddot{\theta} &=& \frac{\xi^2 \cot \theta \csc^2 \theta \dot{\theta}
-a^2 \sin (2\theta) \dot{\theta} /2 -\Sigma \left\{2r \dot{r}-a^2
\sin(2\theta) \dot{\theta} \right\} \dot{\theta}^2}{\dot{\theta}
\Sigma^2} \ . \label{eq:theta2}
\end{eqnarray}
The initial conditions necessary for the integration of the orbit
equations involve, besides the source position, ($r_s,\theta_s)$,
also the two angles of photon directions and the values of $\dot r$
and $\dot \theta$; the last four quantities are not independent but
they can be expressed in terms of ($r_s,\theta_s)$ and the values of
the impact parameters $\eta$ and $\xi$ as discussed in \S 2.3.

\subsection{Locally Isotropic Emission}

In the absence of any information concerning the angular
distribution of the photons emitted at a given flare, the most
conservative assumption is that the emission is isotropic in the
rotating fluid rest frame. However, the proximity of the X-ray
sources to the black hole and the associated source motion ($v_s
\simeq 0.6 \, c$) and strong field geometry along with the fact that
the coordinate system for the integration of the photon orbit
equations is that of the LNRF, require that the notion of photon
isotropy be transformed to this frame and that the impact of the
Kerr metric on the notion of isotropy be considered with some care.
Below we discuss our prescription of local isotropic emission, while
more details can be found in Appendix A.

\begin{figure}[t]
\epsscale{0.5} \plotone{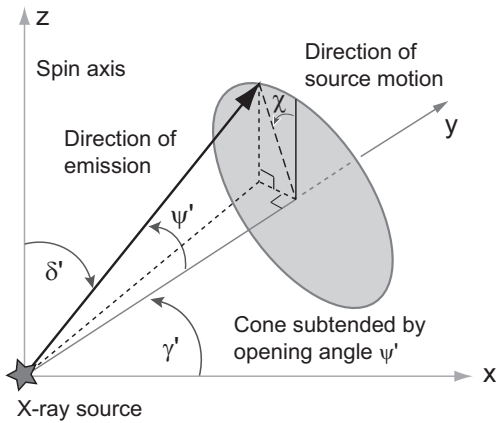} \caption{Geometry of local emission
in the LNRF described in the text. \label{fig:angle}}
\end{figure}

The geometry of photon emission is depicted in
Figure~\ref{fig:angle}. We assume the black hole spin axis to be in
the z-direction while the local fluid velocity $v_s$ in the
y-direction; {$\psi$ is the angle between the emitted photon and the
instantaneous direction of source motion (y-axis) in the fluid
frame, while $\psi'$ is the corresponding angle in the LNRF. For a
given $\psi$ the precise direction of emission is determined by the
angle $\chi$ confined in the plane orthogonal to the y-axis.}
%
%
The relation between $\psi$ and $\psi^{\prime}$ is then (the angle
$\chi$ remains invariant)
\begin{equation}
\psi'(\beta,\psi) = \cos^{-1} \left(\frac{\beta+\cos \psi}{1+\beta
\cos \psi}\right) \ , \label{eq:boost}
\end{equation}
where $\beta \equiv v_s / c$, while the differential solid angle and
opening angle are transformed respectively as
\begin{equation}
d (\cos \psi') = \frac{1-\beta^2}{(1+\beta \cos \psi)^2} \, d (\cos
\psi) \ , ~~~~~~~~ d \psi'(\beta,\psi) =
\frac{(1-\beta^2)^{1/2}}{1+\beta \cos \psi} d \psi \ .
\label{eq:boost2}
\end{equation}

Isotropy in the fluid frame implies equal number of photons per unit
solid angle $N_0$, i.e. $dN/d\Omega \equiv N_0$ over the entire sky
in the fluid rest frame. However, when computing photon geodesics we
choose incrementally not the solid angle but the angles $\psi$ and
$\chi$. Then, the number of photons emitted per increment of the
polar angle $\psi$ is $dN/d \psi = 2 \pi N_0 \sin \psi$ and
therefore the increment of the angle $\chi$ should be equal to $1/(2
\pi N_0 \sin \psi)$ so that  each solid angle element at a given
$\psi, \chi$ contains one photon.

\begin{figure}[t]
\epsscale{1} \plottwo{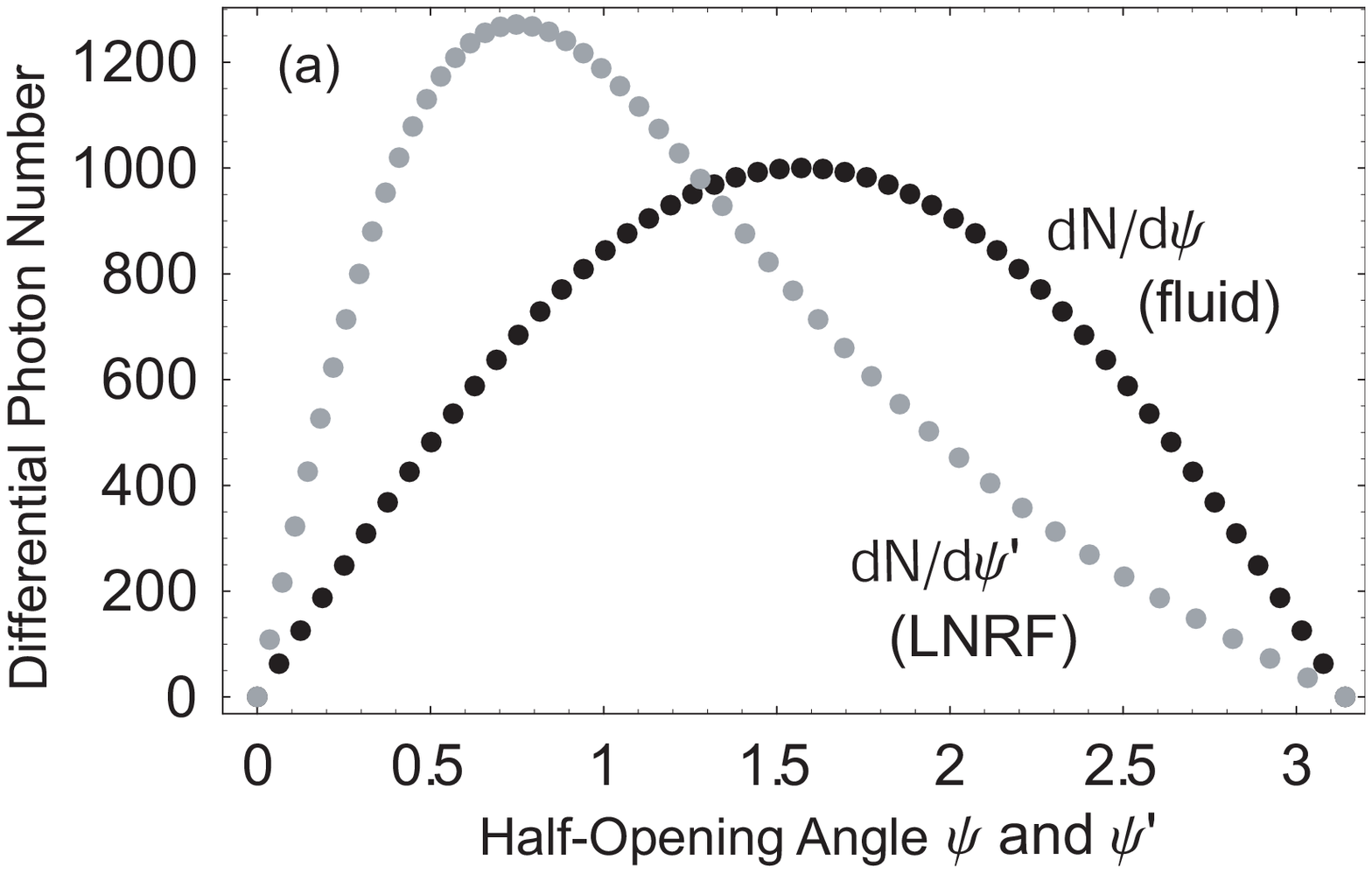}{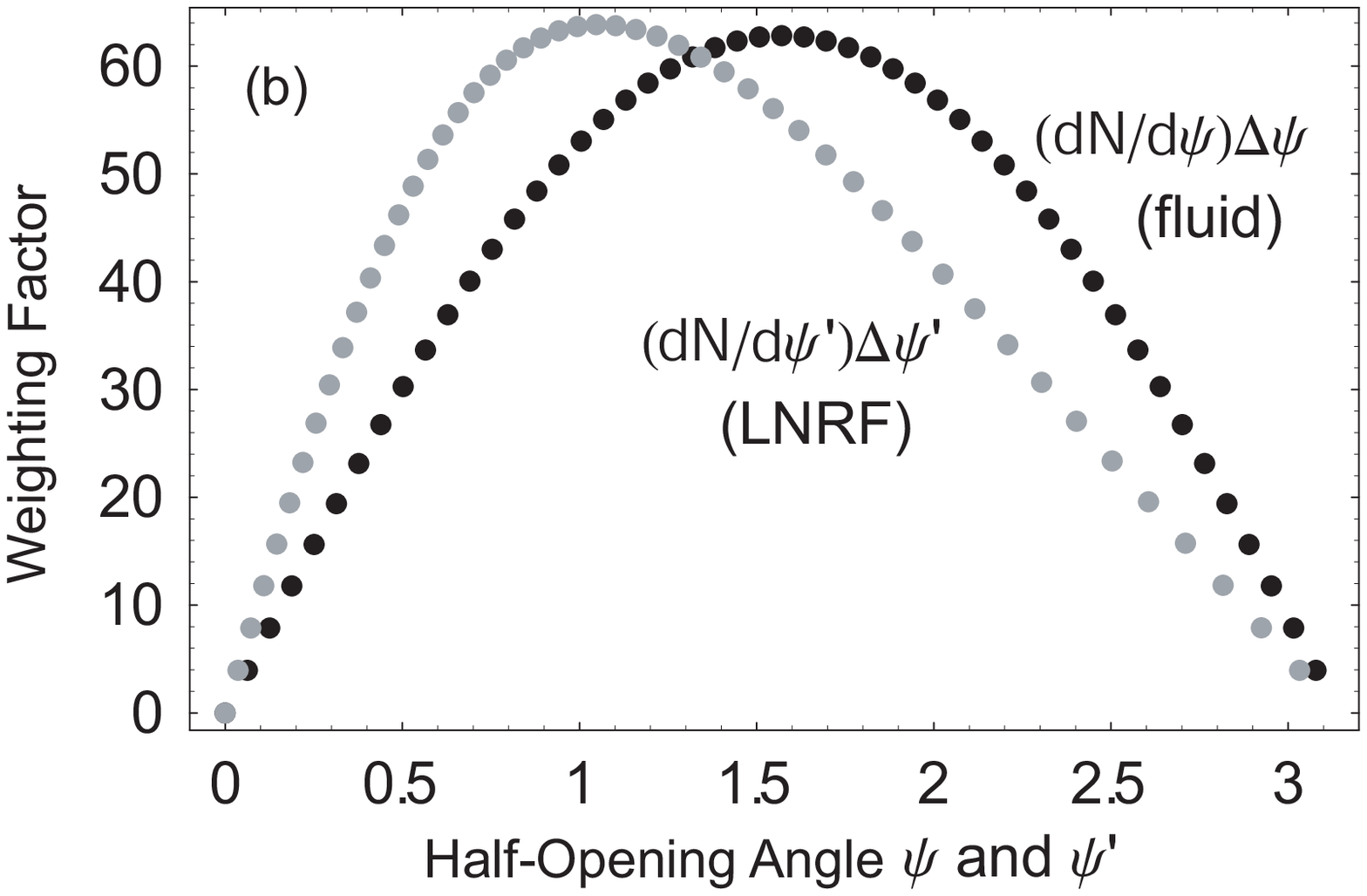} \caption{Comparison between
the fluid rest-frame and the LNRF with $\beta=0.5$ for (a)
Differential photon number distribution per emitting angle as a
function of respective emitting angle in the fluid frame (dark) and
the LNRF (gray). (b) Corresponding bin-corrected photon distribution
(i.e. weighting factor) in the fluid frame (dark) and the LNRF
(gray). We take $N_0=1000$ and $\Delta \psi =\pi/50$.
\label{fig:weighting}}
\end{figure}

The Lorentz boost to the LNRF skews the photon distribution with angle
resulting in larger number of photons being emitted along the direction
of source motion with the distribution as a function of the polar angle
$\psi$ (or $\psi'$) given by the function
\begin{equation}
f(\beta,\psi') \equiv \frac{(1-\beta^2) \sin \psi'}{(1-\beta \cos
\psi')^2} =  \frac{(1+\beta \cos \psi) \sin \psi}{(1-\beta^2)^{1/2}}
\equiv f(\beta,\psi) \ . \label{eq:f}
\end{equation}
In Figure~\ref{fig:weighting}a the photon distribution as a function
of the polar angle $\psi$ for $\beta =0$ (dark dots for the fluid
frame) and also as a function of $\psi'$ for $\beta = 0.5$ (gray
dots for the LNRF) is shown. We take $N_0 =1000$ and $\Delta \psi =
\pi/50$. However, as indicated above in equation~(\ref{eq:boost2}),
the Lorentz boost transforms also the width of the the angular bin
$\Delta \psi$; because of that, as shown in Appendix A, the number
of photons per angular bin is invariant (as it should for photon
number conservation), proportional to $\sin \psi \Delta \psi$. So
each angular bin $\Delta \psi'$ in the LNRF has the same number of
photons as the corresponding bin $\Delta \psi$ in the fluid rest
frame, but these photons are now received at a smaller angle
$\psi'$, given in terms of $\psi$ by equation~(\ref{eq:boost}). This
distribution is shown in Figure~\ref{fig:weighting}b, where the
{number of photons per angular bin} $f(\beta, \psi')\Delta \psi' =
(dN/d \psi') \Delta \psi' = (dN/d \psi) \Delta \psi$ -- for the same
values of $N_0$ and $\Delta \psi$ as in Figure~\ref{fig:weighting}a
-- is shown as a function of the angle $\psi$ (dark dots for the
fluid frame) and $\psi'$ (gray dots for the LNRF). It is apparent
that a given number of photons per bin is found at smaller angles in
the LNRF rather than in the fluid rest frame.

For our photon orbit calculations we transform an equally spaced
(discrete) set of angles $\psi_i$ in the source rest frame to the
corresponding angles $\psi'_i$ in the LNRF and release at each such
bin a number of photons equal to $2 \pi N_0 \sin \psi_i$. These
photons are then distributed each in a single of as many equal
intervals in the $\chi$ angle. The next step involves expressing
local emitting angles $\delta'$ and $\gamma'$ (both measured in the
LNRF) used in the integration of the photon trajectories in terms of
the angles $\psi'$ and $\chi$. The relations between these sets of
angles can be obtained from trigonometric relations of the geometry
shown in Figure~\ref{fig:angle} as
\begin{equation}
\cos \psi' = \sin \delta' \, \sin \gamma' \ , ~~~~~~ \cos \delta' =
\sin \psi' \cos \chi \ , \label{eq:angles}
\end{equation}
where the first of these equations was obtained from a spherical
triangle for the cosine of the angle $\psi'$ while the second one
from relations of the projection of the geometry of the same figure
onto the plane perpendicular to the y-axis (i.e. in the shaded
ellipse of the same figure).

The next step in the calculation takes into account the effects of
the local geometry curvature in the definition of local isotropy
\citep[see][as an example of this issue in the simpler case of the
\sw geometry]{FK07} by relating the angles $\delta'$ and $\gamma'$
to the photon four-velocity components. In the LNRF the photon's
local emitting polar and azimuthal angles ($\delta',\gamma'$) are
related to the four-velocity components through the relations
\begin{eqnarray}
v^{\hat{r}} &=& \frac{A^{1/2}}{\Delta} \frac{\dot{r}}{\dot{t}} =
\sin \delta' \cos \gamma' \ , \label{eq:vrhat}
\\
v^{\hat{\theta}} &=& \left(\frac{A}{\Delta}\right)^{1/2}
\frac{\dot{\theta}}{\dot{t}} = \cos \delta' \ , \label{eq:vthetahat}
\\
v^{\hat{\phi}} &=& \frac{A \sin \theta}{\Sigma \Delta^{1/2}}
\left(\frac{\dot{\phi}}{\dot{t}}-\omega\right) = \sin \delta' \sin
\gamma' \ , \label{eq:vphihat}
\end{eqnarray}
which lead to
\begin{eqnarray}
\cos \delta' &=& \left(\frac{A}{\Delta}\right)^{1/2}
\frac{\dot{\theta}}{\dot{t}} \ , \label{eq:eqn1}
\\
\cot \gamma' &=& \frac{\Sigma}{A^{1/2} \Delta^{1/2} \sin \theta}
\frac{\dot{r}}{\dot{\phi}-\omega \dot{t}}  \ , \label{eq:eqn2}
\end{eqnarray}
evaluated at the emitting location ($r_s,\theta_s,0$).
%
Using equations~(\ref{eq:t}) through (\ref{eq:phi}) we can eliminate
$\dot{t}$, $\dot{r}$, $\dot{\theta}$ and $\dot{\phi}$ from the above
equations~(\ref{eq:eqn1}) and (\ref{eq:eqn2}) in favor of the two
impact parameters $\eta$ and $\xi$. We are thus led to two relations
that connect the photon emission angles ($\delta',\gamma'$) to the
photon impact parameters $(\xi,\eta)$. In fact, one can invert these
relations to obtain $(\xi,\eta)$ as a function of these angles, i.e.
$\xi=\xi(\delta',\gamma')$ and $\eta=\eta(\delta', \gamma')$ as
shown in Appendix B. This now completes the problem of the choice of
the orbit parameters consistent with isotropic emission in the fluid
frame: For a given set of equally spaced values of $\psi_i$ (locally
isotropic source in the fluid rest frame) one can compute $\psi'_i$
and also determine the number of photons to be emitted and values of
the azimuthal angle bins $\chi_i$; these are then used to compute
the angles $\delta'$ and $\gamma'$, and then the corresponding
impact parameters $\eta$ and $\xi$ along with the initial values of
$\dot r(\delta', \gamma')$ and $\dot \theta(\delta', \gamma')$ that
are used in the integration of the photon orbit by
equations~(\ref{eq:r2}) and (\ref{eq:theta2}).

\begin{figure}[t]
\epsscale{1} \plottwo{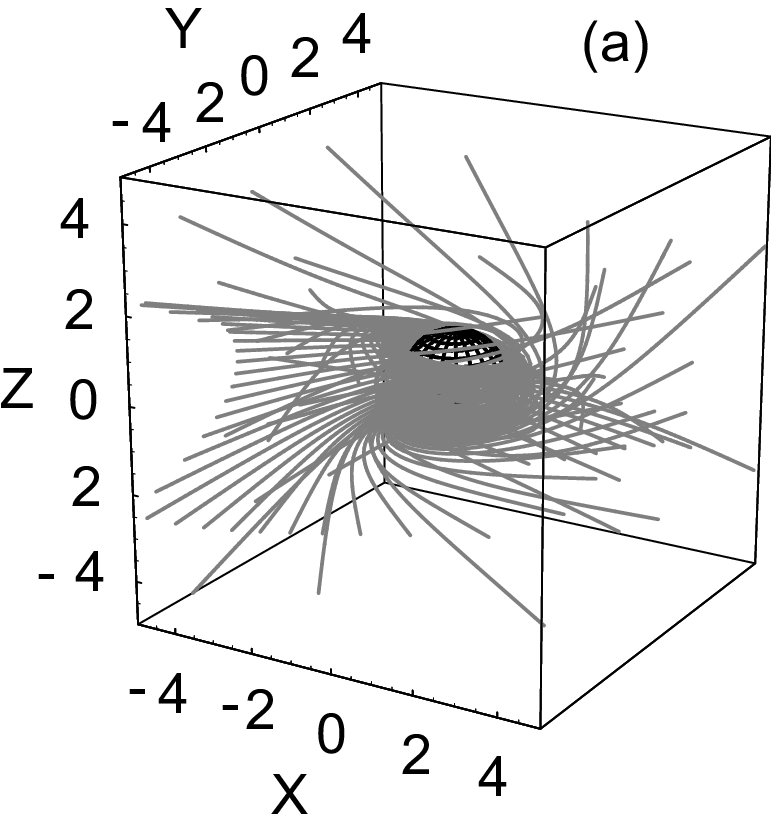}{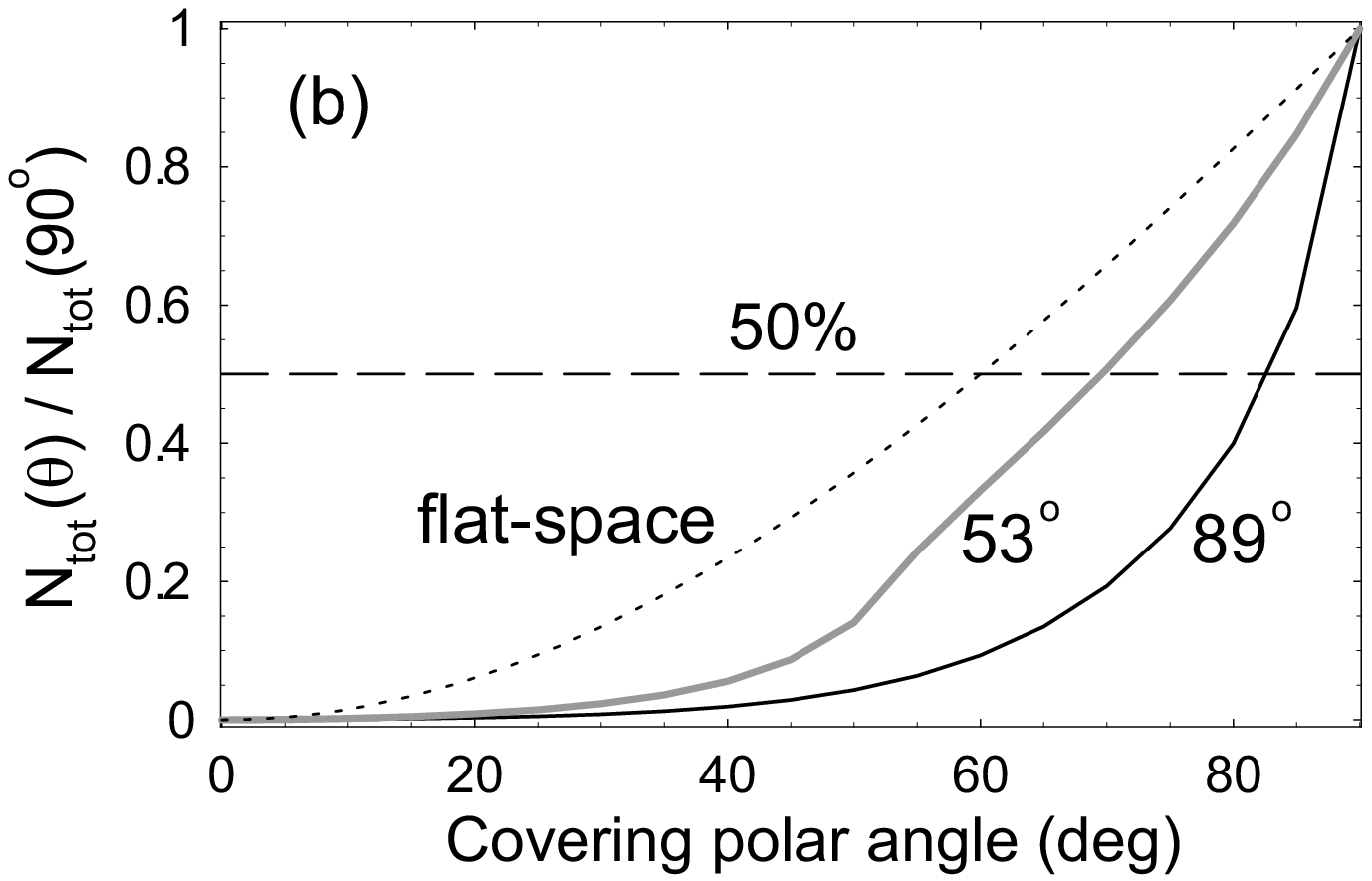} \caption{(a) Locally
isotropic photon trajectories from a source at $(\sin
\theta_s,\phi_s)=(89^\circ,0^\circ)$ for an opening angle of $\psi =
80^\circ$ ($\psi'=46^\circ$). We set $a/M=0.99$ and $N_0=100$. (b)
Observed (cumulative) photon distribution $N_{\rm
tot}(\theta)/N_{\rm tot}(90^\circ)$ as a function of covering angle
measured from the pole for various source geometries. It is
normalized by the total photon number over $\pi/2$. We assume an
equatorial source of $\theta_s=89^\circ$ (dark curve) and
mid-latitude source of $53^\circ$ (gray curve), in comparison with
stationary flat-space case (dotted curve). \label{fig:ray}}
\end{figure}

An example of 3D photon trajectories from a locally isotropic
emission is shown in Figure~\ref{fig:ray}a where we consider
emission with a half-opening angle of $\psi=80^\circ$
($\psi'=46^\circ)$ from a source at
$(\theta_s,\phi_s)=(89^\circ,0^\circ)$. Throughout this paper we
assume $a/M=0.99$ as in Figure~\ref{fig:ray}a, unless otherwise
stated. Recall that our model considers now X-ray sources from both
the upper and lower hemispheres with respect to the equator because
of the optically-thin assumption of surrounding matter (see
Fig.~\ref{fig:geometry}). Hence, the net observed emission will
consist of a sum of these two components. We have confirmed that
these photon trajectories are indeed almost symmetric with respect
to the equator as expected. {It should be noted, however, that for
some particular emission angles this symmetry in trajectory with
respect to the equator can be largely broken even for a small
deviation of the source position from the equator.} Due to the high
velocity of the source ($\beta \sim 0.6$) for the values of $r_s$
and $a$ chosen, most of the emission is beamed forward in the
direction of instantaneous source velocity. In fact we find that
many backward-emitted photons (i.e. of $\psi>90^\circ$) subject to
the Doppler and frame-dragging effects of the fast black hole
rotation are forced to corotate in the hole direction (see FK08 for
details in 2D cases). We also find that backward-emitted photons
with $\psi \gtrsim 140^\circ$ end up crossing the the horizon,
making no contribution to the observed signal. We should mention
that the photon trajectories considered here also include many
multiple orbits and higher-order photons; i.e. we also collect those
photons that orbit around a black hole multiple times with some
time-lags. The physical significance of these time-lags will be
further discussed in \S 2.4.

To illustrate the effects of Doppler beaming and the black hole's
strong gravity for the considered source location ($r_s \sin
\theta_s \simeq 1.45 \, M$), we present in Figure~\ref{fig:ray}b the
cumulative, normalized distribution of photons $N_{\rm
tot}(\theta)/N_{\rm tot}(90^\circ)$ received by a distant observer
at $r/M=100$, as a function of the observer's inclination angle
$\theta_o$. In the figure the solid dark curve corresponds to a
nearly equatorial source ($\theta_s = 89^{\circ}$), the gray curve
to a source at the edge of the ergosphere ($\theta_s \simeq
53^{\circ}$) for the fixed cylindrical radius of the source, while
the dotted curve to the flat-space distribution. It is apparent in
this figure that photons emitted from equatorial position (in
particular) are directed primarily at low latitude directions. While
in flat-space 50\% of the photons (horizontal dotted line) are
emitted between angles $0^{\circ} - 60^{\circ}$ as expected, in the
equatorial source case of $\theta_s=89^\circ$ the half-way point has
been extended to $\theta \simeq 82^{\circ}$. Similarly, while in
flat-space polar angles $0^{\circ} - 45^{\circ}$ include $\simeq
30\%$ of the entire photons emitted, in the present Kerr geometry
they include roughly $10$ times less. This is an important issue
related to the equivalent width of the (broad) fluorescence Fe lines
whose red wing is presumably emitted by plasma closest to the ISCO
of fast-rotating black holes with spin parameter similar to that
used in producing our Figure~\ref{fig:ray}b.

\subsection{Response Functions and Light Curves}

Following a methodology similar to that discussed in FK08, we first
compute a source's response function seen by a distant observer
whose radial and azimuthal positions are set in this work to
$(r_o/M,\phi_o) = (100,0)$. This is a reasonable choice because at
this distance the photon orbits are already nearly straight and thus
their final (polar and azimuthal) directions are almost the same as
those in flat spacetime. For a given source position at
($\theta_{s,i},\phi_{s,i}$) around a rapidly-rotating black hole we
collect isotropically emitted photons (in the local rest-frame of an
orbiting source) keeping track of its arrival time $t$ and final
position ($\theta,\phi$) for each photon. As a result we can
construct a response function seen by the observer
$I_i(t,\theta_o;\theta_{s,i},\phi_{s,i})$, which tells us photon
counts as a function of photon's arrival time from the i-th X-ray
source at ($\theta_{s,i},\phi_{s,i}$). For an optically-thin
environment we have contributions from sources at different source
latitudes for the same source azimuth, however, for simplicity we
assume that the sources closer to the equatorial plane are dominant,
although it is found that this assumption does not make much
difference to our final results shown \S 3 below.

\begin{figure}[t]
\epsscale{1} \plottwo{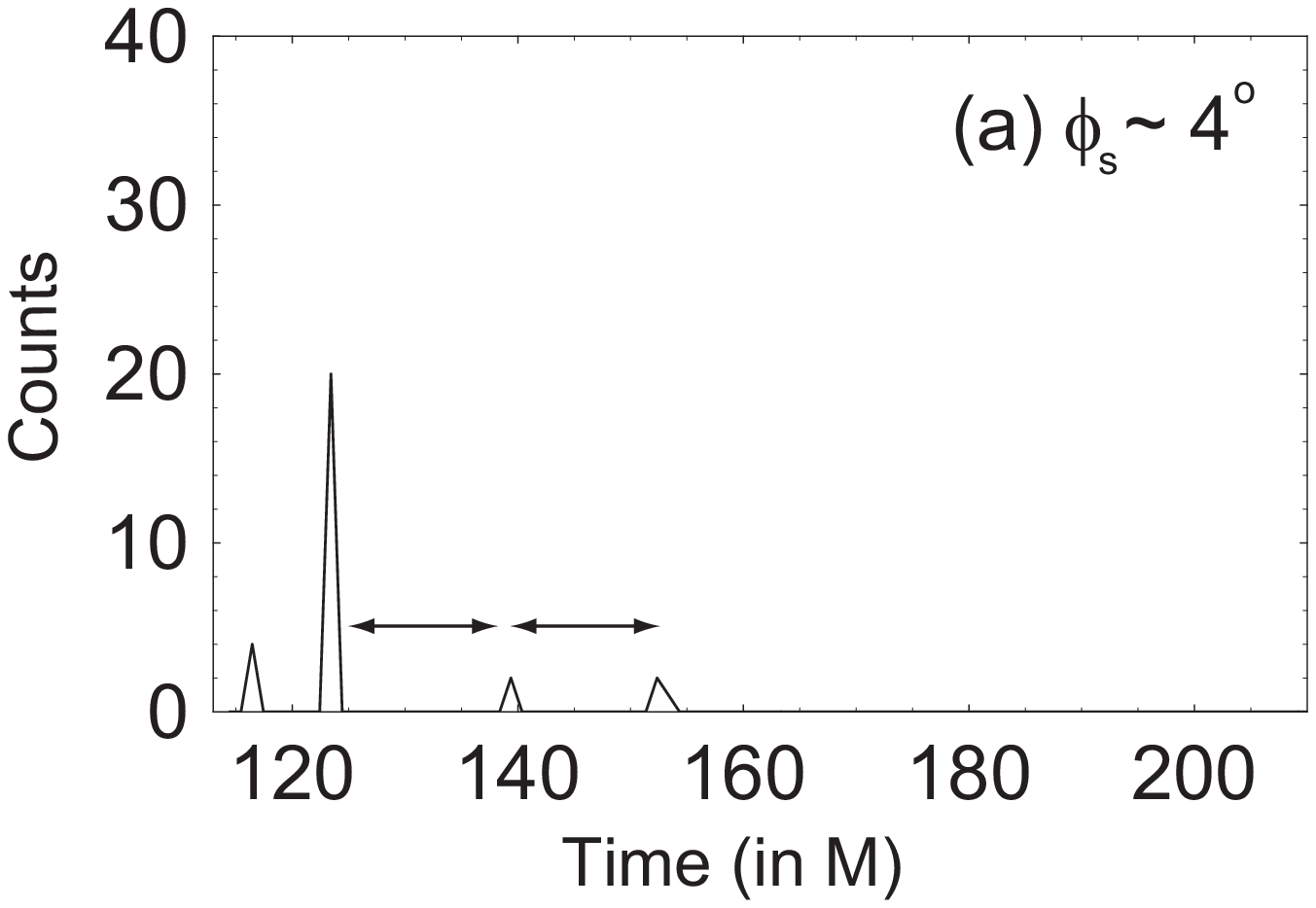}{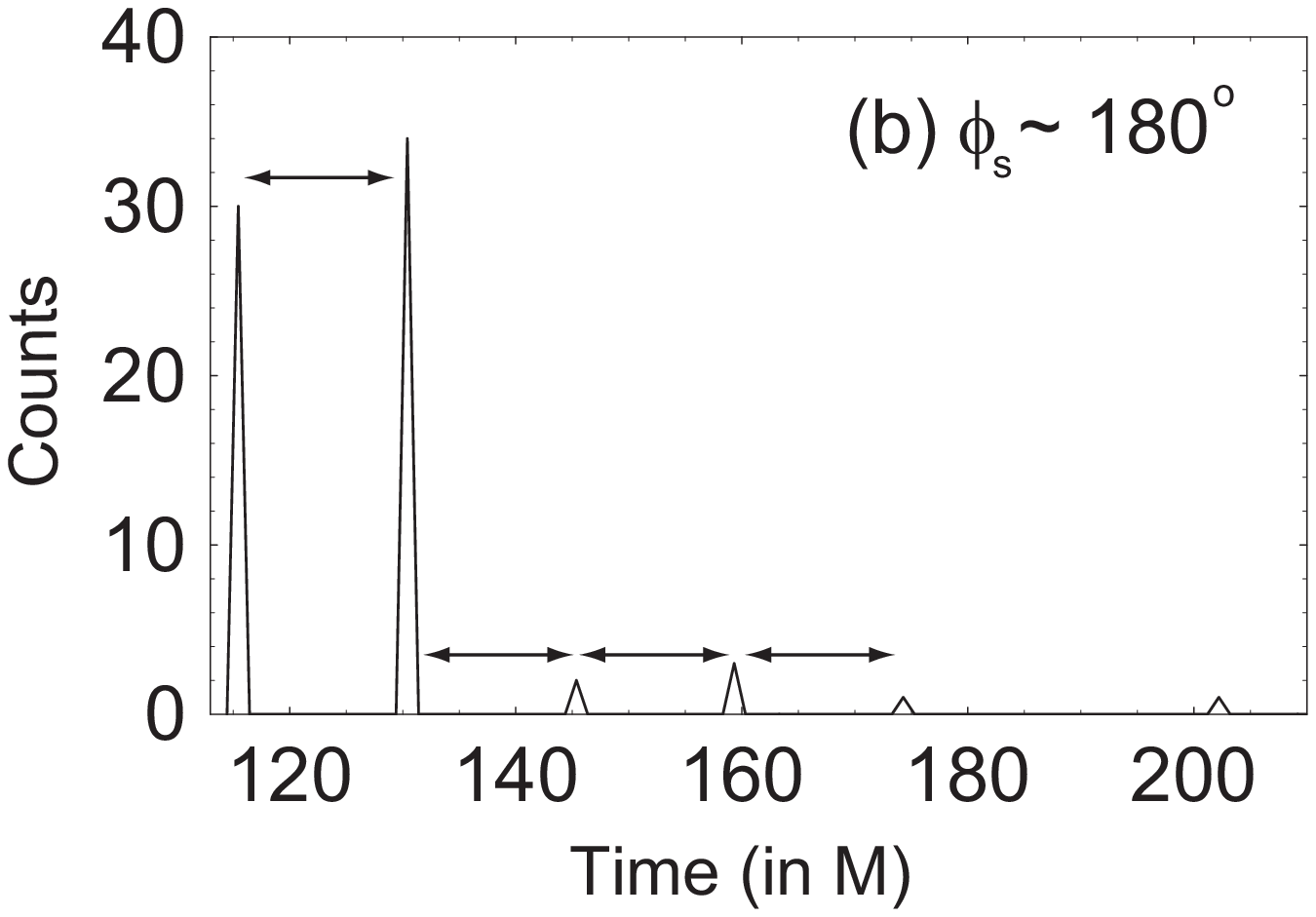} \caption{Synthetic response
functions from a single X-ray source at $\theta_s=89^\circ$ with (a)
$\phi_{s} \sim 4^\circ$ and (b) $\sim 180^\circ$ seen by an observer
of $\theta_o = 60^\circ$. Arrows indicate a constant lag of $\sim
14M$. Values of the other parameters are the same as in
Figure~\ref{fig:ray}. \label{fig:response}}
\end{figure}

The temporal resolution in the response function is taken to be $1M$
(corresponding to $5 \times 10^{-5}$ sec for a $10\Msun$ black
hole), therefore the signal received by the observer is uniformly
binned with this numerical accuracy. For computational purposes we
approximate this signal, as precisely as possible, by a set of
narrow Gaussians, also of width $1M$. Note the response function
from each single source contains a characteristic temporal profile
due to distinct photon trajectories between the source and the
observer, and the profile also strongly depends on the observer's
polar position $\theta_o$ as well as the source position
$(\theta_{s,i},\phi_{s,i})$. The observer's collecting angle is
chosen to be $\Delta \theta_{o} = \pm 1^\circ$. To produce a
statistically significant outcome even for observer's polar angles
$\theta_o \lsim 30^{\circ}$ we consider a large number of sampling
(a few $10^6$) per source. However, the common feature they share is
found to be a constant time separation of $\simeq 14 \, M$ between
the major peaks of individual response functions, as found in FK08.
To clearly illustrate these features of the response functions, we
exhibit in Figure~\ref{fig:response} the response functions from
flares at (a) $\phi_s \sim 4^\circ$ and (b) $\sim 180^\circ$ for
$\theta_s=89^\circ$ measured by an observer of $\theta_o=60^\circ$.
A constant time lag of $\sim 14 M$ (equivalent to $\sim 0.7$ msec
for a $10 \Msun$) is clearly present. Due to multiple orbits around
a fast-rotating black hole some photons make extra
(integer-multiples) full orbits before reaching the observer.

%
%

The response functions are different in the precise positions of the
major peaks, as one would expect, but, as argued above, the time-lag
between the major peaks is roughly constant, $\Delta t \sim 14 \, M
$, regardless of the source location ($\theta_s,\phi_s$). The reason
for this behavior was discussed in detail in FK08 (it remains the
same in the present case too): it is due to the frame-dragging
effects which force all photons emitted at any relative position
between source and the observer, to reach the latter only by
``swinging" around the black hole; in following such trajectories, a
non-negligible fraction ($\simeq 15-20\%$) of them reach the
observer after multiple orbits around the black hole, inducing
``light echoes" of nearly constant lag as shown in
Figure~\ref{fig:response}. This is in agreement with similar
relativistic calculations of broad iron line profiles by
\citet{Beckwith05} where they find that these higher-order photons
contribute maximally $\sim 20-60$\% of the total luminosity of the
system. The persistent presence of such a constant time-lag in the
response functions is the key factor leading to QPO features in the
power spectra shown in \S 3. It should be noted that in the absence
of strong frame-dragging (i.e. sources outside the ergosphere or \sw
instead of Kerr black hole), the time interval among the peaks in
the response functions is phase-dependent and precludes the presence
of QPO in the power spectra for random source positions (see FK08).

As discussed in \S 2.1 the X-ray interval between flares $\Delta T$
is Poisson-distributed with mean value comparable to the Keplerian
period at the ISCO. Therefore, the collective response function from
the ensemble of X-ray flares (i.e. the bolometric light curve) is
obtained by summing over every contribution from all the sources at
$(\theta_{s,i},\phi_{s,i})$. That is,
\begin{eqnarray}
I(t,\theta_o) \equiv \sum_{i=1}^{n}
I_i(t,\theta_o;\theta_{s,i},\phi_{s,i}) \ , \label{eq:lc}
\end{eqnarray}
where $n$ is a total number of random X-ray sources in our
simulations. As an example, a synthetic light curve is shown in
Figure~\ref{fig:lc}a (i.e. superposition of all the response
functions similar to Fig.~\ref{fig:response}) and its first $20 M_1$
msec segment in Figure~\ref{fig:lc}b where $M_1 \equiv M/(10\Msun)$.
In this specific case, the number of flares considered is $n=2000$
with the source positions confined to near the equatorial plane
($\theta_{s}=89^\circ$) while the observer is set at $\theta_o=
60^\circ$. Because of the random positions ($\phi_{s,i}$) and
bursting times of individual flares, the model light curve appears
to be (in fact it is constructed to be) totally random with no
apparent coherence (even in the absence of background noise).
Nonetheless, the characteristic time-lag among the peaks in flux
implicit in the form of the response function in
Figure~\ref{fig:response} is encoded in the light curve and it can
be extracted (if present) by applying the standard time-series
analysis techniques discussed in the next section \S 3.

\begin{figure}[t]
\epsscale{1} \plottwo{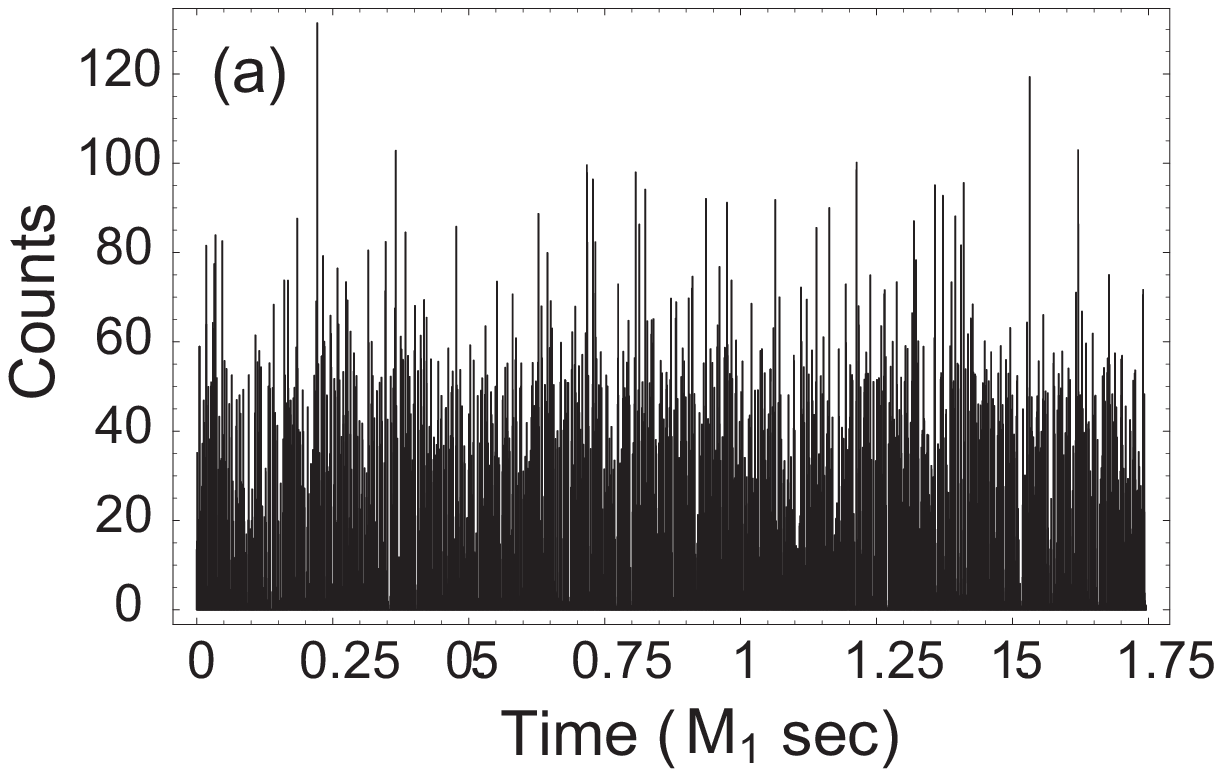}{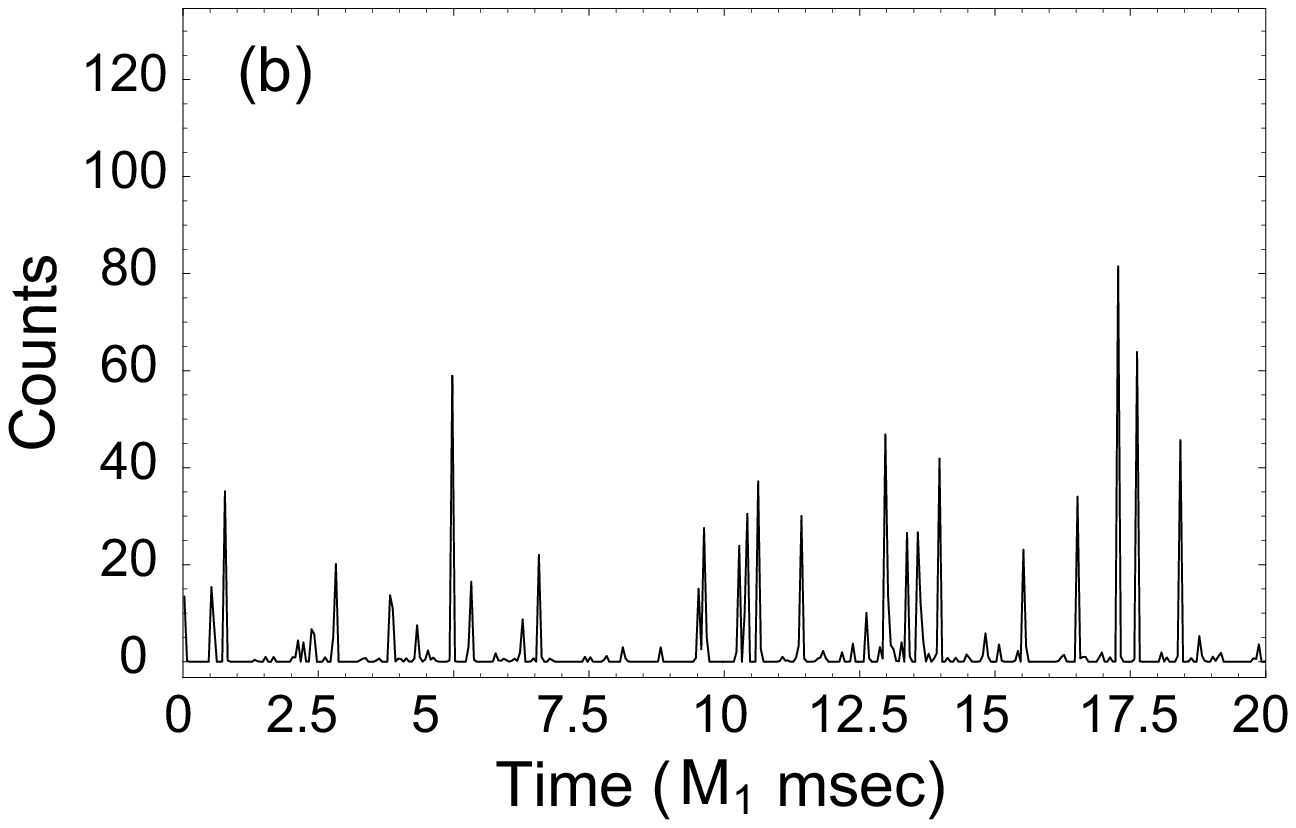} \caption{(a) Model light
curves constructed from the synthetic response functions similar to
those in Figure~\ref{fig:response}. We assume $n=2000$ random X-ray
sources from $\theta_{s}=89^\circ$ for $\theta_o=60^\circ$. (b) The
first $20 M_1$ msec segment of (a). Values of the other parameters
are the same as in Figure~\ref{fig:ray}. \label{fig:lc}}
\end{figure}

\section{Power Spectra and Autocorrelation Functions}

In this section we discuss in detail the timing analysis metrics,
ACFs and PSDs, of the model light curves produced using the
prescriptions discussed above, pertaining to rapidly-rotating black
holes; more specifically we study the dependence of the ACFs and
PSDs on the sources' position and geometric structure as well as the
position of the observer.

As shown in \S 2 above, the form of the response functions and hence
of the model light curves we produce depend mainly on the sources'
latitudes (given that we average over their azimuths $\phi_{s,i}$).
Because the effects of frame-dragging are most prominent for sources
near the equator, we first examine cases where most of the X-ray
emission is concentrated preferentially near the equatorial plane,
i.e. of $\theta_{s,i}=89^\circ$ (and corresponding positions in the
lower hemisphere). Regarding the observer's inclination angle
$\theta_{o}$ we consider a wide range from nearly face-on
($\theta_{o} \simeq 30^\circ$) positions to nearly edge-on
($\theta_{o} \simeq 85^\circ$) ones. Since our primary goal in this
investigation is to generalize the previous 2D results of FK08, we
restrict ourselves to a fast-rotating black hole case with
$a/M=0.99$ for which the source X-ray photons are subject to strong
dragging of inertial frame (i.e. inside the ergosphere). Note that
for $a/M=0.99$ the ISCO radius is naturally well inside the
ergosphere near the equatorial plane.

We also consider, for a given observer's angle $\theta_{o}$, the
dependence of the ACF and PSD on the latitudinal distribution of the
sources within the ergosphere. To this end we compute also the
response function for source positions at different vertical heights
with $\Delta \theta_{s} = 9^\circ$ by dividing the polar angle into
five representative zones; i.e. $\theta_s = 89^\circ$, $80^\circ$,
$71^\circ$, $62^\circ$, and $53^\circ$ (see
Fig.~\ref{fig:geometry}b). We then examine the effects of the
sources' vertical extent by computing light curves from sources with
incrementally larger vertical extent, i.e. from sources whose
emission consists of the sum of point sources at a number of heights
corresponding to the following set of inclination angles:
$\theta_{s}= \theta_{89^\circ}$ (i.e. equatorial sources),
$\theta_{89^\circ+80^\circ}$, $\theta_{89^\circ+80^\circ+71^\circ}$,
$\theta_{89^\circ+80^\circ+71^\circ+62^\circ}$, and
$\theta_{89^\circ+80^\circ+71^\circ+62^\circ+53^\circ}$. We do not
consider sources at heights that would correspond to angles smaller
than $53^\circ$, because for the given value of black hole spin
($a/M=0.99$) and a vertical source whose equatorial foot point is at
the ISCO, a larger height would place that section of the source
outside the ergosphere, thereby producing no contribution to the QPO
features via frame-dragging effect. Given that the effects of frame
dragging decrease with increasing source height we expect the QPO
effects to decrease with increasing vertical source extent.

The timing properties of the model light curves are explored for
different relative positions between the source's and observer's
polar angles by computing the ACF, $R(\tau;\theta_o,\theta_s)$, and
the corresponding PSD, $P(\nu;\theta_o,\theta_s)$.
For a discrete light curve $I(t_o, \theta_o; \theta_{s})$ the normalized
ACF is given by
\begin{eqnarray}
R(\tau_j;\theta_o,\theta_s) &\equiv&  \sum_{k=1}^{L}
\frac{I(t_k,\theta_o;\theta_{s}) I(t_{k+j},\theta_o;\theta_{s})} {
I^2(t_k,\theta_o;\theta_{s})} \ , \label{eq:R}
\end{eqnarray}
where $L$ is the number of time bins in the light curve of $I$. The
corresponding PSD, in units of (rms/mean)$^2$/Hz, is given by
\begin{eqnarray}
P (\nu_j;\theta_o,\theta_s) &\equiv& \frac{2T}{N_{\rm ph}^2} |a_j|^2
\ , \label{eq:FourierAmp}
\end{eqnarray}
where $T$ is the time duration of the light curve $I$ and $a_j$ is
the Fourier amplitude defined by
\begin{eqnarray}
I(t_k,\theta_o;\theta_{s}) = \frac{1}{L} \sum_{j=1}^{L} a_j e^{2\pi
i j k/L} \ , \label{eq:Fourier}
\end{eqnarray}
and $N_{\rm ph}$ is the total photon counts in the entire light
curve of $I$. Also, $\tau_j$ and $\nu_j$ are timescale and frequency
of a characteristic QPO signal, respectively.

\subsection{Dependence on Viewing Angle }

In this subsection we first present in Figure~\ref{fig:obs-dep} a
series of ACFs (left panels) and PSDs (right panels) as a function
of the observer's inclination (or viewing) angle $\theta_{\rm o}$
for model light curves from both near-equatorial sources, i.e. of
$\theta_{s}=89^\circ$, and a source that is vertically extended in
height, modeled as a sum of individual sources at heights that
correspond to polar angles $\theta_s = 89^{\circ}$, $80^{\circ}$,
and $71^{\circ}$. These are shown in Figures~\ref{fig:obs-dep}a and
b (equatorial source) and c and d (vertically extended source).

\begin{figure}[t]
\centering
\begin{tabular}{cc}
\epsfig{file=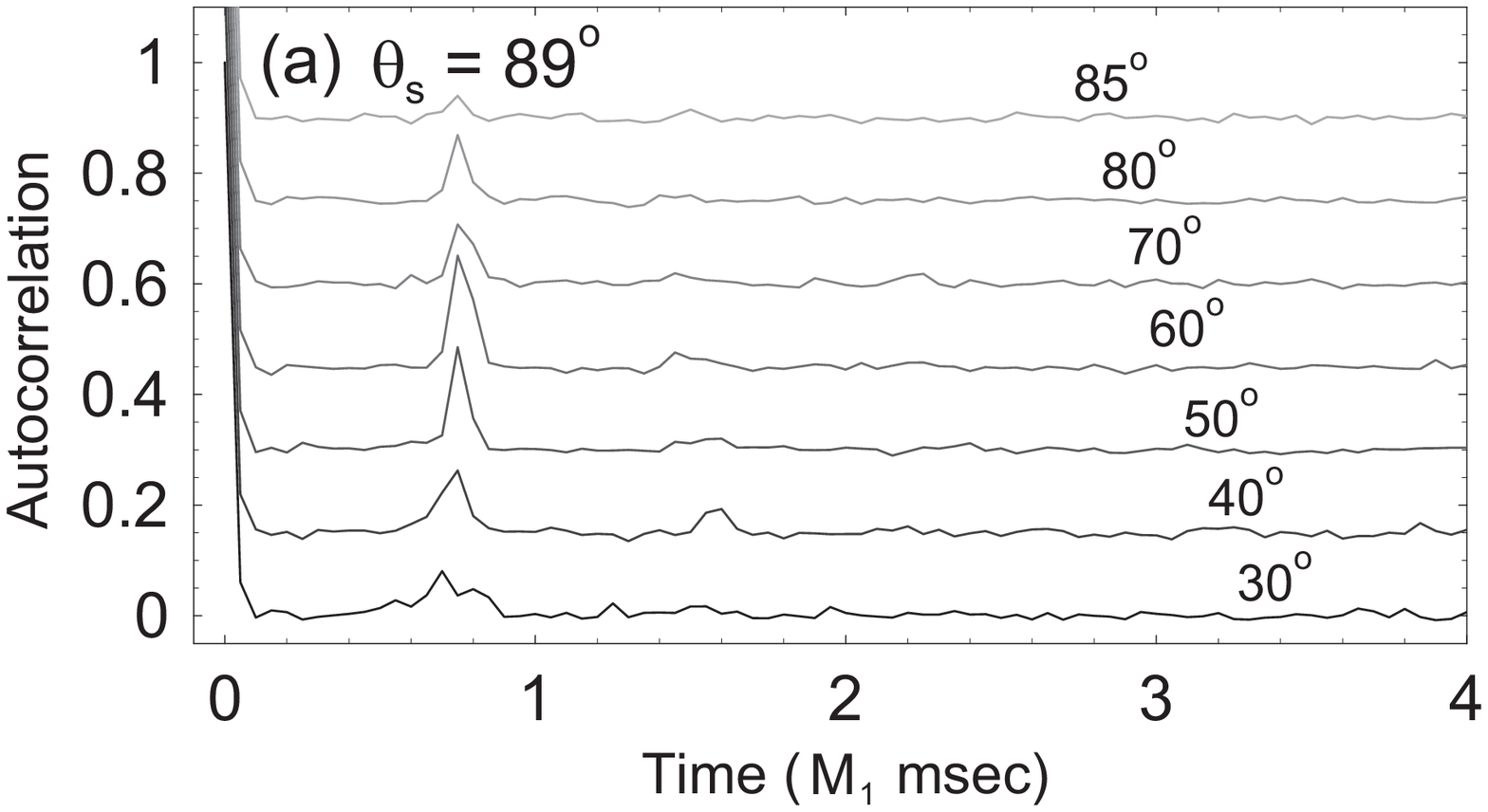,width=0.45\linewidth,clip=} &
\epsfig{file=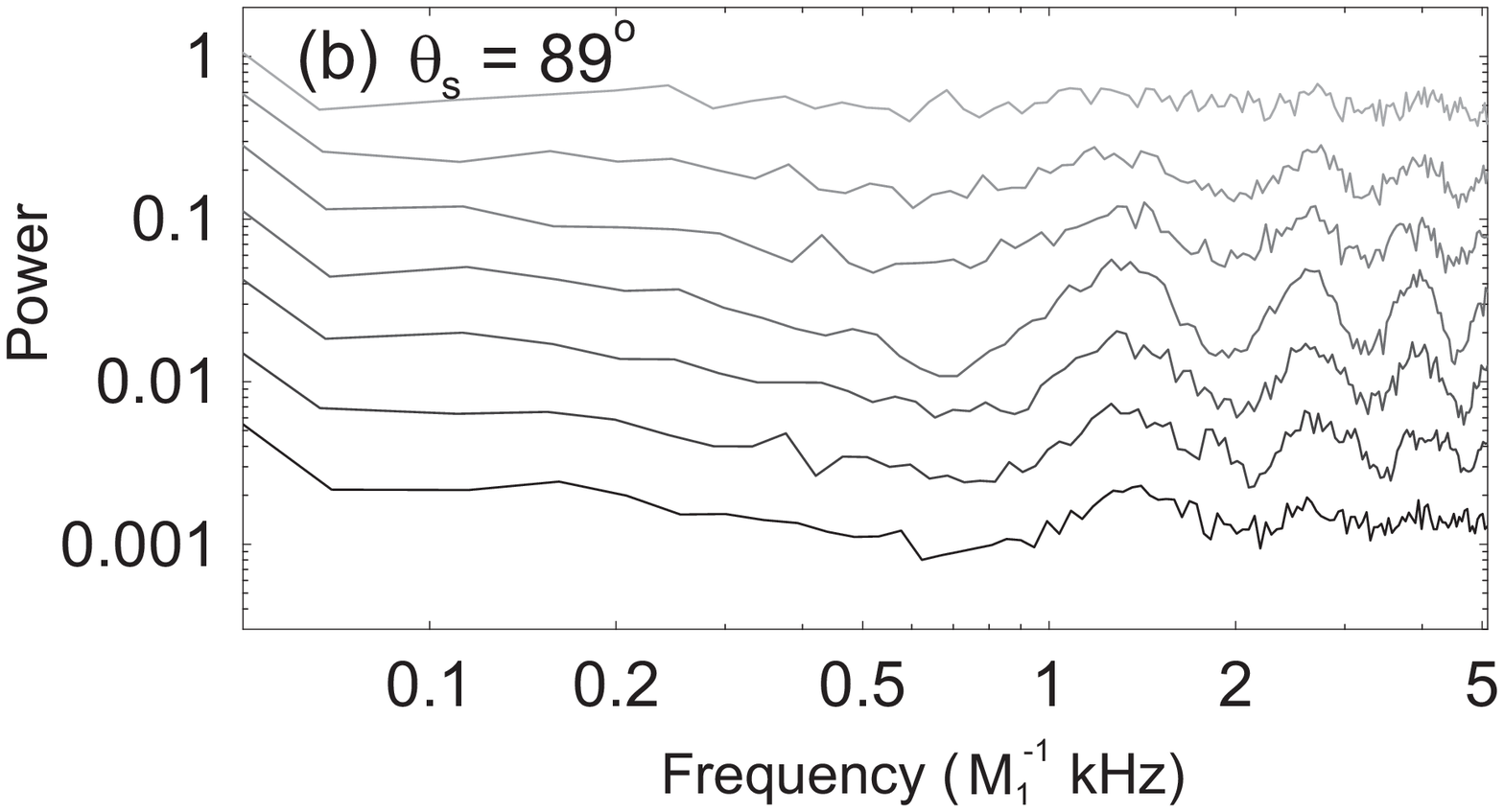,width=0.45\linewidth,clip=} \\
\epsfig{file=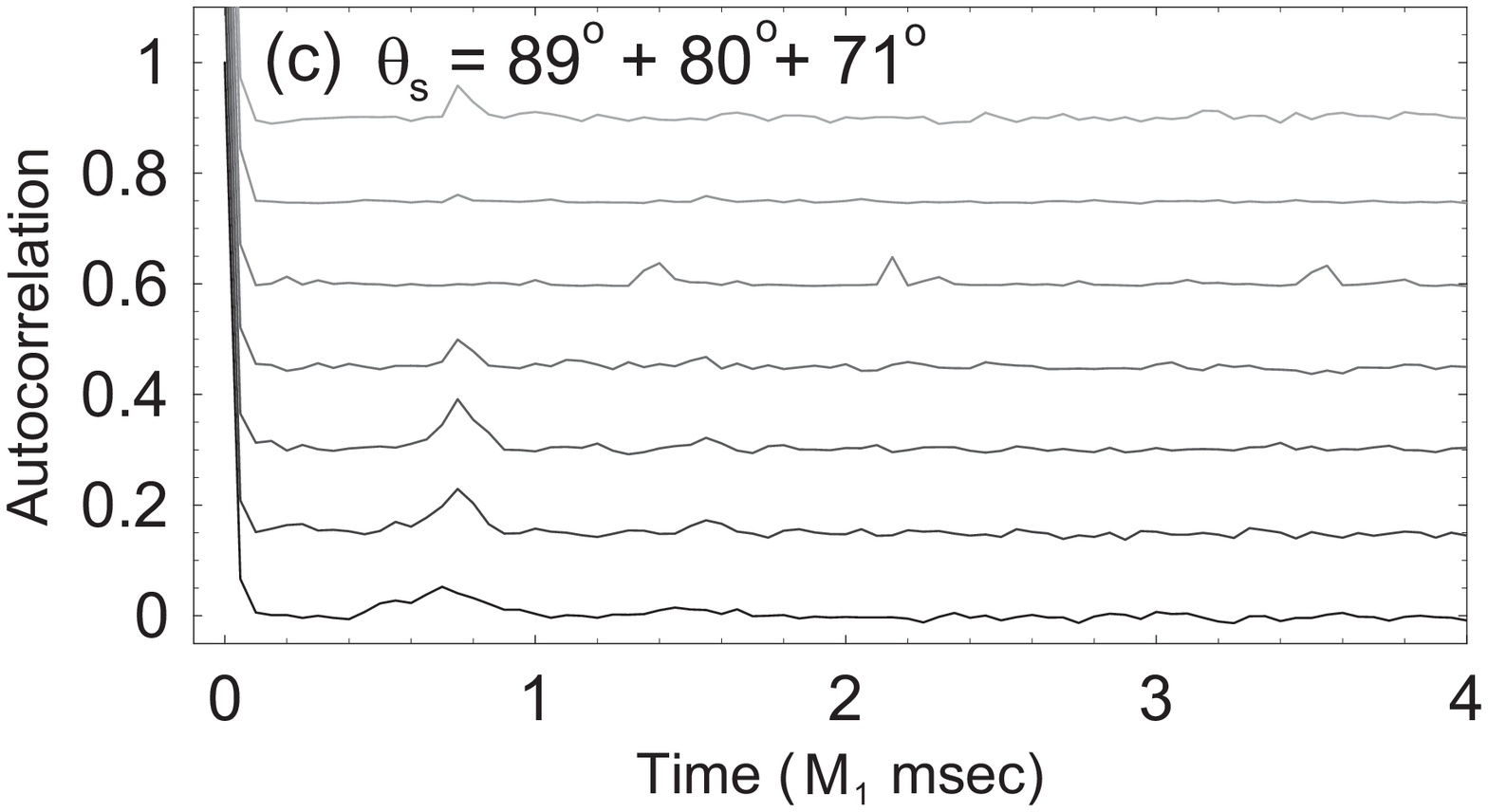,width=0.45\linewidth,clip=} &
\epsfig{file=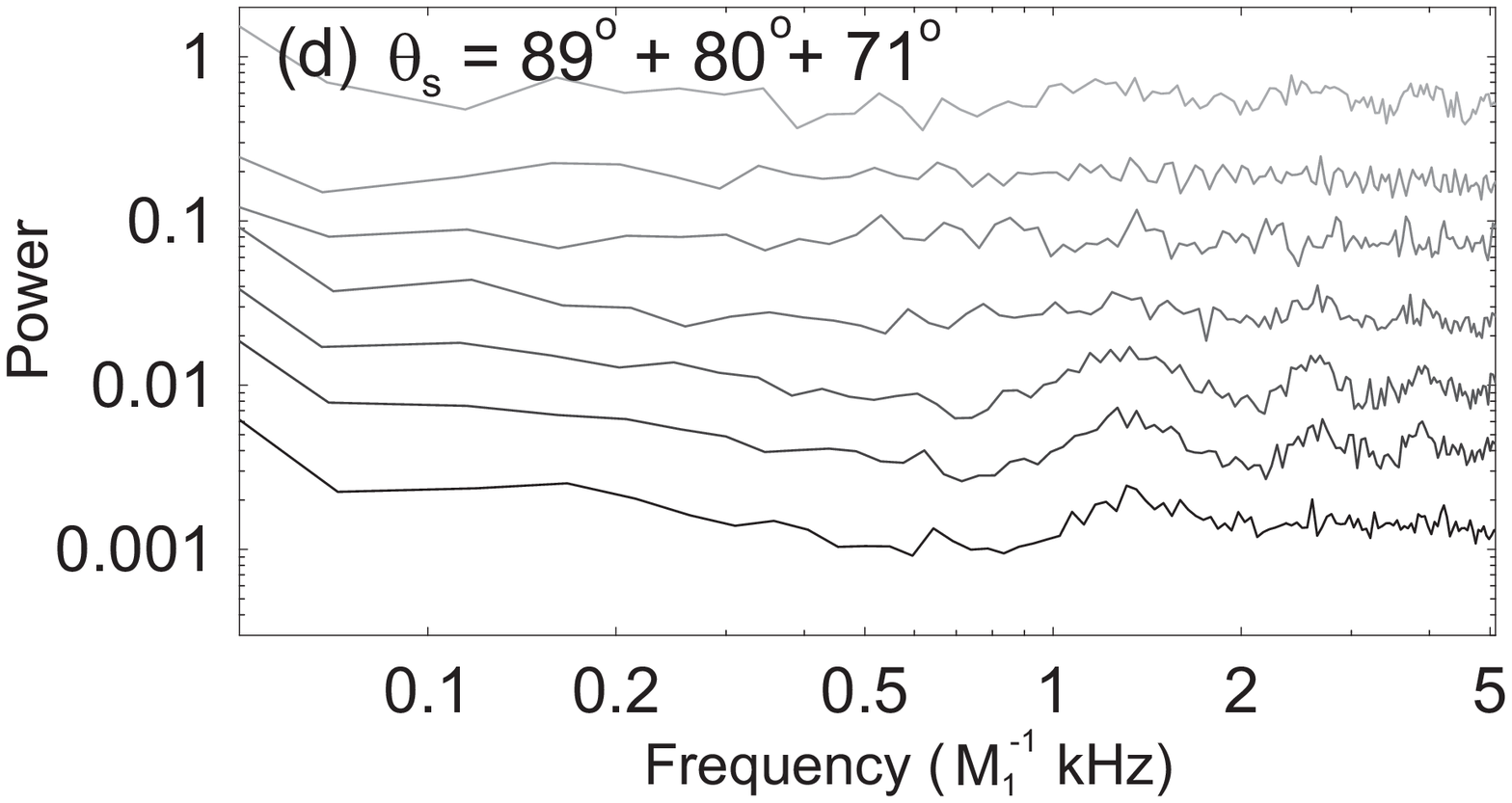,width=0.45\linewidth,clip=}
\end{tabular}
\caption{Autocorrelation functions (left panels) and power spectral
densities (right panels) for various observer's inclination angles
where $\theta_s=\theta_{89^\circ}$ (upper panels) and
$\theta_s=\theta_{89^\circ+80^\circ+71^\circ}$ (lower panels). We
show
$\theta_o=30^\circ,40^\circ,50^\circ,60^\circ,70^\circ,80^\circ,85^\circ$
from bottom to top curves with $\Delta \nu = 0.6$ Hz bin size.
Vertical positions of the curves are arbitrarily shifted for
presentation purpose. Values of the other parameters are the same as
in Figure~\ref{fig:ray}. \label{fig:obs-dep}}
\end{figure}

As noted in FK08 the QPO signal in the light curves is imprinted by
the frame dragging effects induced by the black hole spin. {It is
seen that a characteristic timescale of $\tau \sim 14M =0.73 M_1$
msec is persistently present for a wide range of inclination
$\theta_{o}$, which is a manifestation of the constant time-lag in
the response function in Figure~\ref{fig:response}.} Since these
effects are more prominent for photon orbits that lie near the
equator, this signature is more pronounced for arrangements that
maximize the propagation of the received photons in this region. As
shown in Figures~\ref{fig:obs-dep}a and b this effect is the most
optimized for $\theta_{\rm o} \sim 50^\circ - 60^{\circ}$; for
observers at smaller inclination (polar) angles the effect is weaker
because the influence of frame-dragging is much reduced for photons
that propagate close to the black hole spin axis. This becomes
apparent by the decrease of the ACF peak at $\tau \simeq 0.73 M_1$
msec, or equivalently, the corresponding QPO amplitude with the
decrease in $\theta_{\rm o}$ (Fig.~\ref{fig:obs-dep}a and b). The
QPO amplitude decreases also for $\theta_{\rm o}> 60^\circ$; the
reason for that is the increase of the fraction of the poloidal
plane orbits that connect the source and the observer; these photons
do not orbit around the black hole with coherence and as such they
do not contribute to the QPO amplitude but in fact they dilute it.
That is, the total signal is the result of a superposition of a
mixture of both coherent and incoherent signals.

Similarly, as shown in Figures~\ref{fig:obs-dep}c and d, a
vertically extended source leads to reduction in the corresponding
amplitudes of the ACF and QPO since the frame-dragging effects
decrease as the source approaches the edge of the ergosphere.
Comparison of Figures~\ref{fig:obs-dep}a and b with
Figures~\ref{fig:obs-dep}c and d, which exhibit respectively the
ACFs and PSDs for a set of observers at inclinations $\theta_{\rm
o}= 30^{\circ}$, $40^{\circ}$, $50^{\circ}$, $60^{\circ}$,
$70^{\circ}$, $80^{\circ}$, $85^{\circ}$ for an equatorial (a,b) and
a vertically extended source (c,d), confirms these notions; the QPOs
are more prominent for sources confined near the equator than for
those which are vertically extended. The former also persists over a
wider range of values of $\theta_{o}$, being strongest for
$\theta_{o} \sim 50^{\circ} - 60^{\circ}$, while the latter, being
intrinsically weaker, they disappear for $\theta_{o} \gsim
60^{\circ}$.

Finally, while the QPO and ACF secondary peak amplitudes do depend
on the observer inclination, the corresponding QPO frequency and ACF
peaks remain constant respectively equal to $\nu_{QPO} \sim
1.4M_{1}^{-1}$ kHz and $\tau \sim 0.73M_{1}$ msec, a value
representing the length of the photon paths around the ISCO,
becoming considerably wider only for $\theta_{\rm o} \lsim
30^\circ$.

\subsection{Dependence on Source Concentration }

If the X-ray sources originate from processes other than those
associated with the surface of a thin disk (e.g. flaring events due
to magnetic reconnection, polar shocks or wind/jets), then the
source may not be exclusively distributed in an equatorial fashion.
In this subsection we examine in greater depth the effects of
extending the source dimension vertically on the ACFs and QPOs
respectively. To this end we have produced the light curves starting
with an equatorially concentrated source ($\theta_s = 89^{\circ}$)
and consider the light curves from more extended sources by
incorporating into the light curve also photons from sources at
larger heights, or smaller values of the angle $\theta_s$; this is
done incrementally by considering additional sources of the same
intensity at the following set of source polar angles:
$\theta_{89^\circ+80^\circ}$, $\theta_{89^\circ+80^\circ+71^\circ}$,
$\theta_{89^\circ+80^\circ+71^\circ+62^\circ}$, and
$\theta_{89^\circ+80^\circ+71^\circ+62^\circ+53^\circ}$. We do not
consider sources at any larger heights because for the conditions
used here they would be outside the ergosphere and they would not
contribute to the QPO signal by frame-dragging.

\begin{figure}[t]
\centering
\begin{tabular}{cc}
\epsfig{file=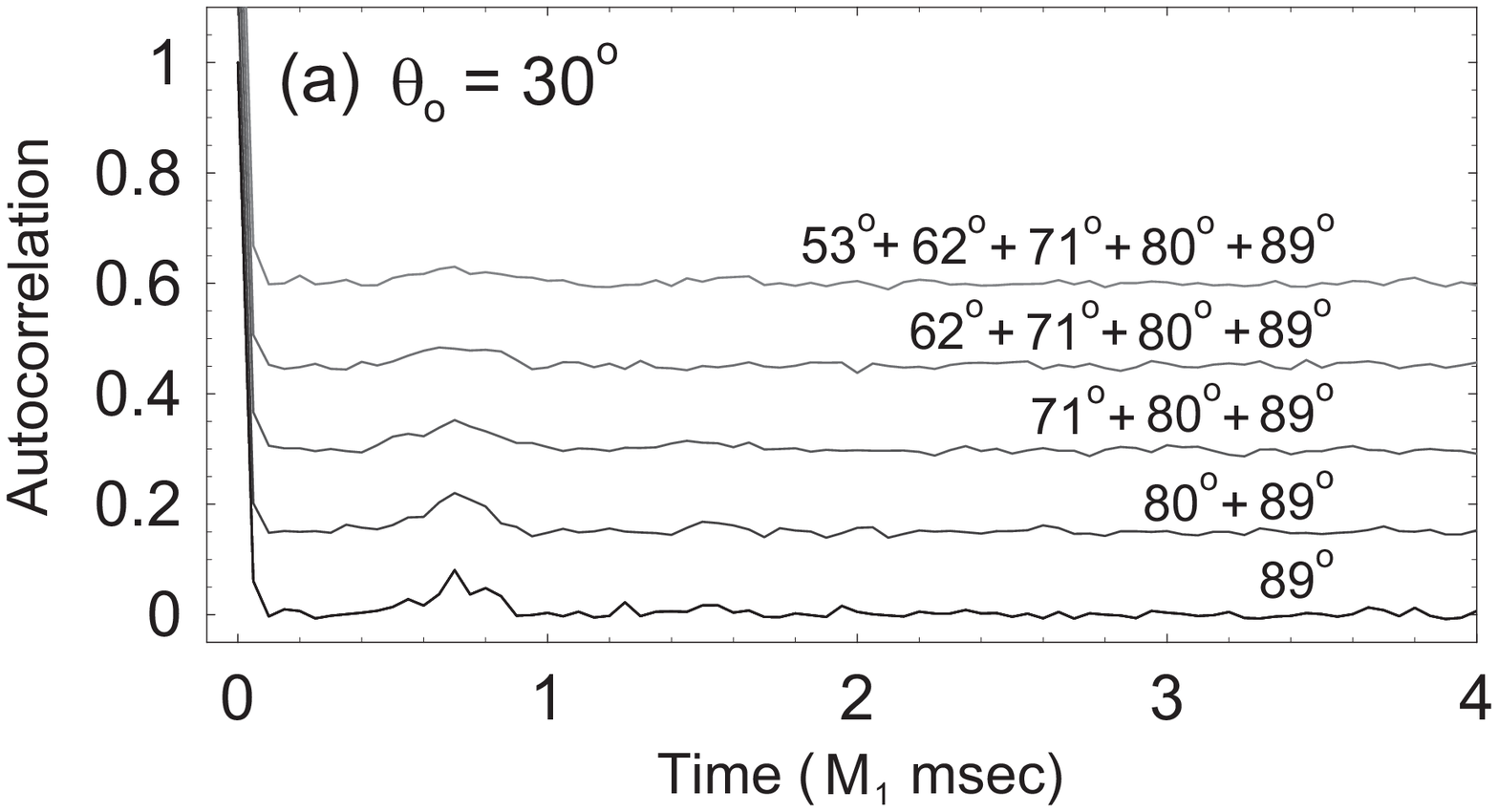,width=0.45\linewidth,clip=} &
\epsfig{file=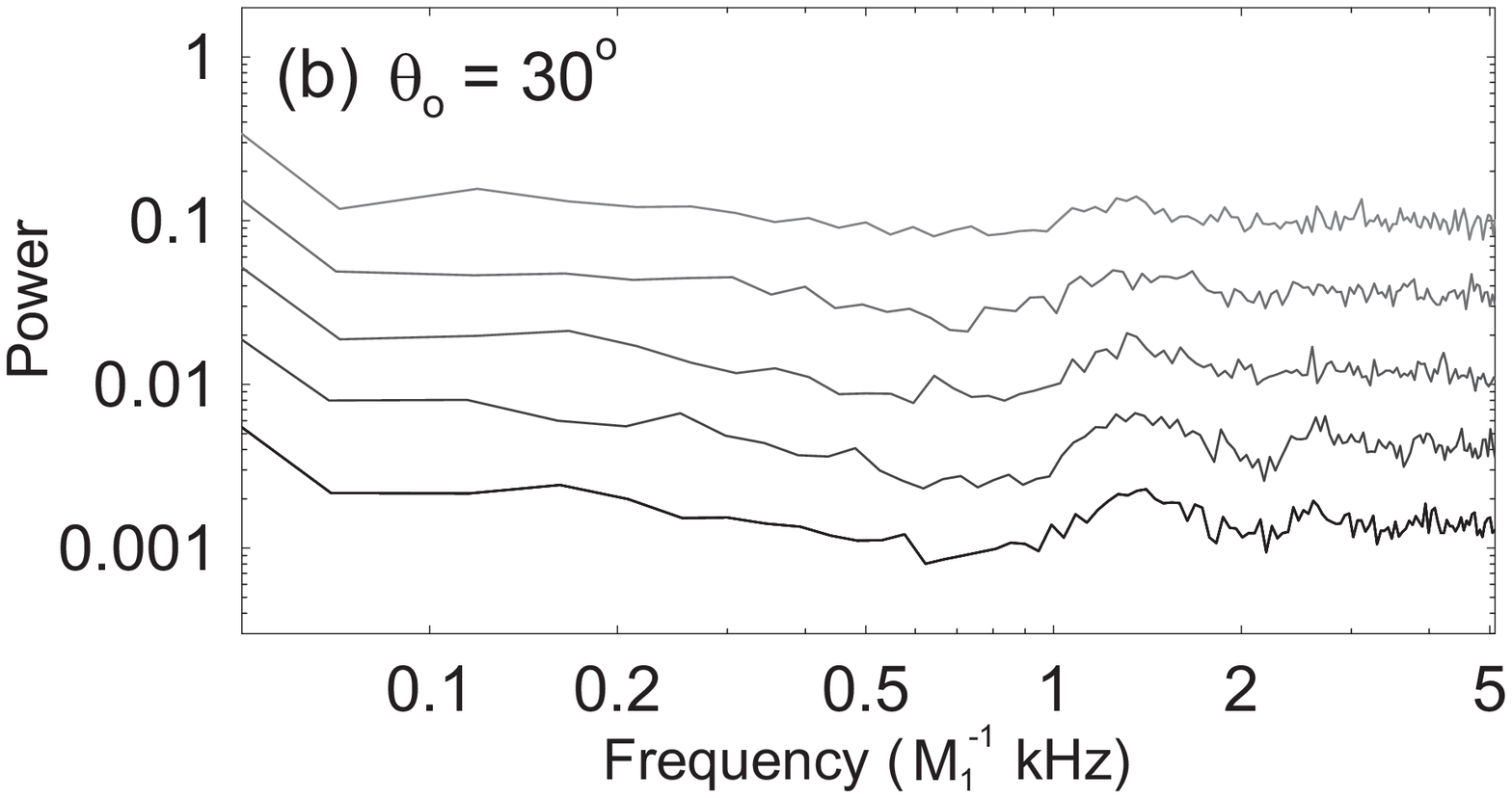,width=0.45\linewidth,clip=} \\
\epsfig{file=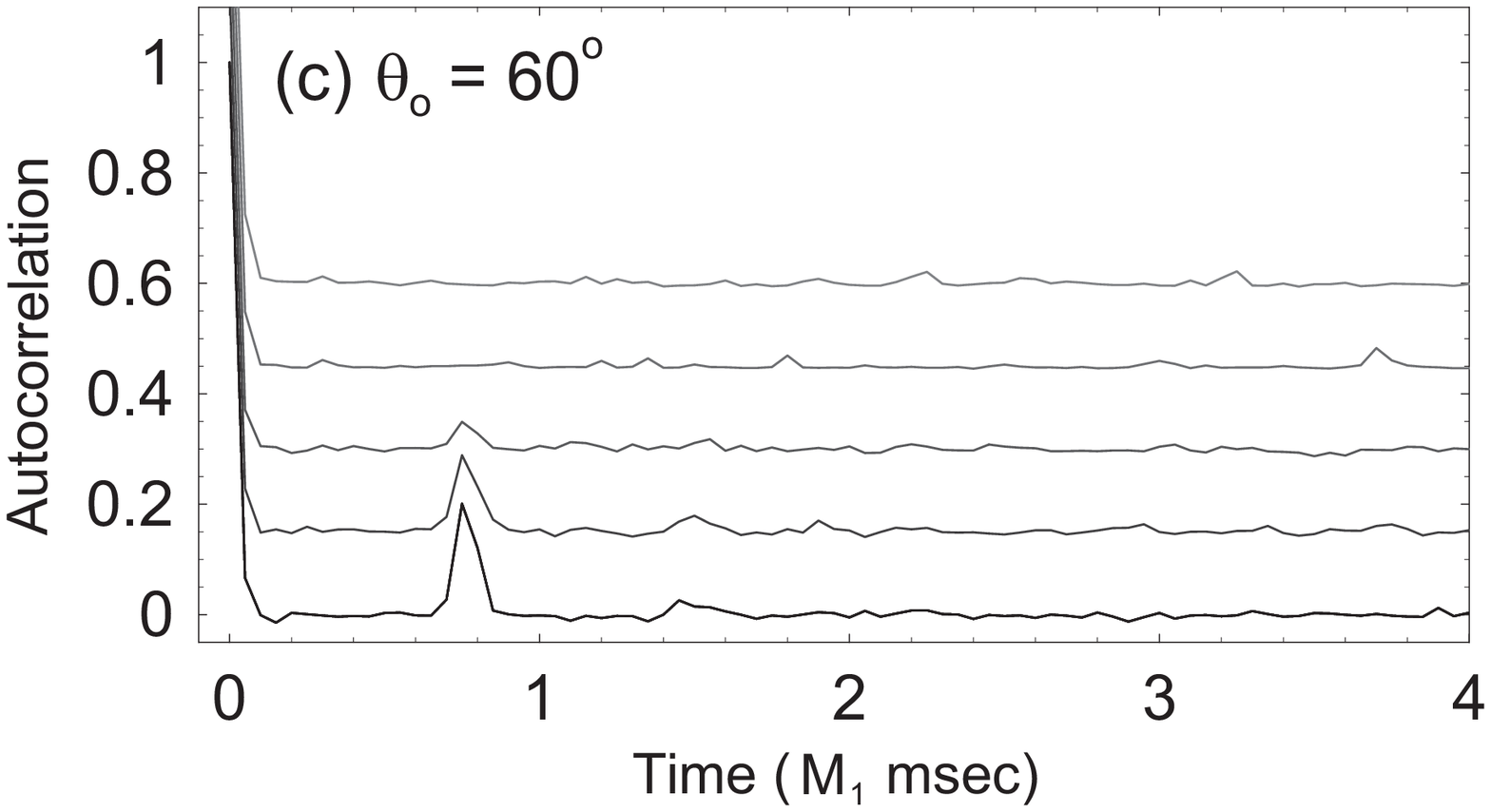,width=0.45\linewidth,clip=} &
\epsfig{file=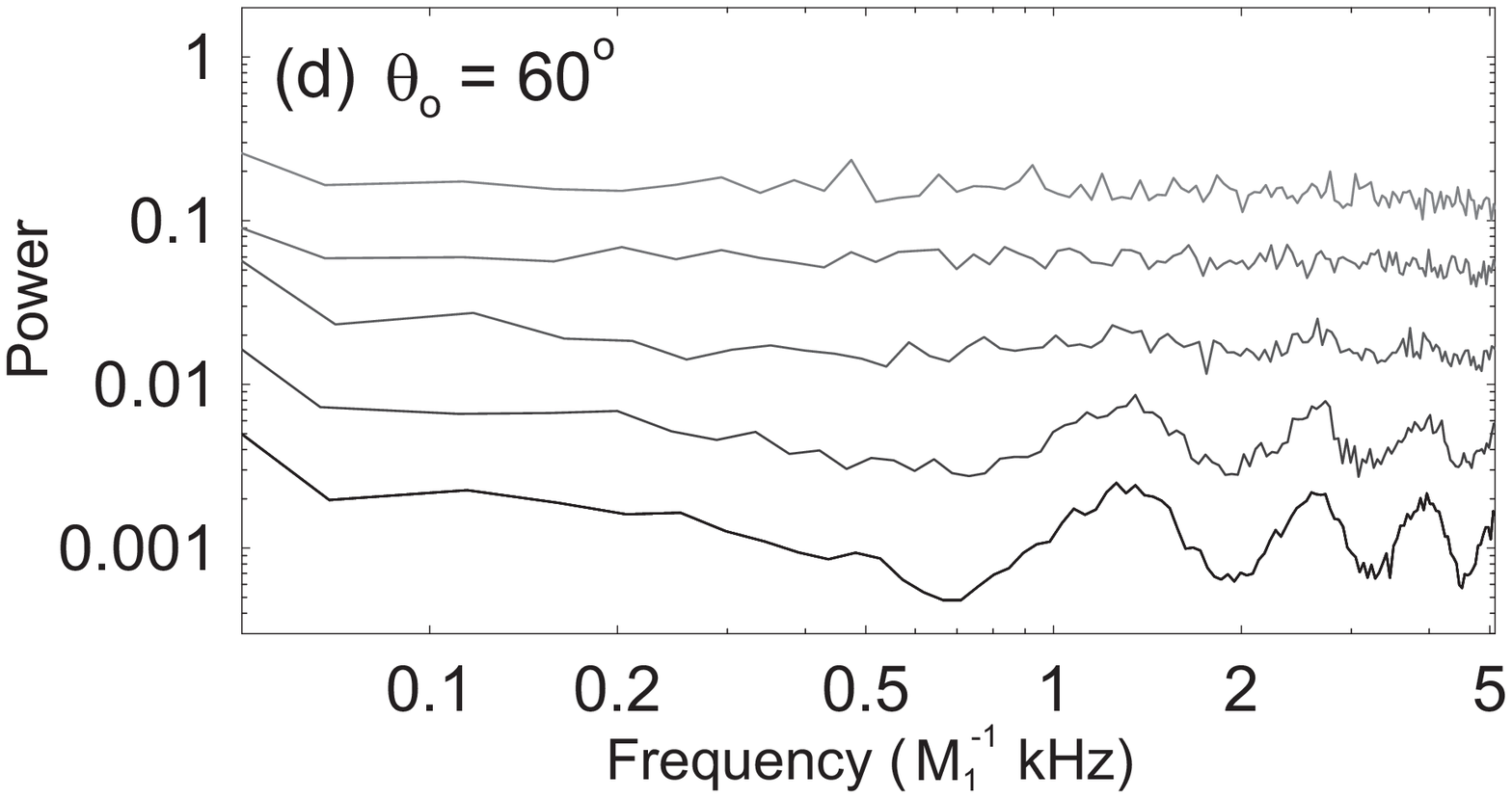,width=0.45\linewidth,clip=}
\end{tabular}
\caption{Autocorrelation functions (left panels) and power spectral
densities (right panels) for various source distributions where
$\theta_o=\theta_{30^\circ}$ (upper panels) and
$\theta_o=\theta_{60^\circ}$ (lower panels). We show $\theta_s =
\theta_{89^\circ}$, $\theta_{89^\circ+80^\circ}$,
$\theta_{89^\circ+80^\circ+71^\circ}$,
$\theta_{89^\circ+80^\circ+71^\circ+62^\circ}$, and
$\theta_{89^\circ+80^\circ+71^\circ+62^\circ+53^\circ}$ from bottom
to top curves with $\Delta \nu = 0.6$ Hz bin size. Vertical
positions of the curves are arbitrarily shifted for presentation
purpose. Values of the other parameters are the same as in
Figure~\ref{fig:ray}. \label{fig:src-dep}}
\end{figure}

In Figures~\ref{fig:src-dep}a and b we present respectively the ACFs
and PSDs for an observer at $\theta_{\rm 0} = 30^\circ$ as the
source height progressively increases to $\theta_s = 53^{\circ}$
(i.e. sources extending all the way up to the static limit).
{Overall, the QPO signal does not appear to be very strong but it is
present.} We see that the additional (incoherent) photons dilute
those with constant lag to cause the QPO to disappear with the
source height. Similarly, in Figures~\ref{fig:src-dep}c and d we
present the ACFs and PSDs for an observer with high inclination of
$\theta_{\rm o} = 60^{\circ}$. We see that in this case the QPO
becomes again prominent when the X-ray sources are distributed near
the equator as the photons that reach these directions are under a
strong influence of frame-dragging. It appears that the latitudinal
extension of the sources up to $\theta_s \lesssim 71^\circ$ can
yield a discernible QPO signature, although it is relatively weak.

\section{Summary and Discussion}

In the present work we have extended the earlier 2D treatment of
FK08 to consider the 3D geometry of photon emission on the timing
properties of accretion flows in the vicinity of rapidly spinning
black holes. This has allowed us to explore the effects of placing
both observers and sources at
latitudes much different from the equatorial ones considered in FK08
and, using these results, to consider also photon sources which are
extended, i.e. sources whose vertical dimension is comparable or
larger than their radial one.
Our results are fundamentally consistent with those of
FK08: we find that for the same reasons discussed in FK08, here too,
frame-dragging causes the photons to follow trajectories that by
and large lead to a significant fraction of them reaching the
observer after an additional orbit around the black hole; as such,
for a single X-ray flare, an observer sees multiple bunches of
photons that arrive with a constant time-lag of order of $\sim 14M$
(i.e. {\it light echoes}), independent of the source position
relative to the observer. It is the independence of the lag on the
source position that leads to the QPO in the signal of even a random
ensemble of X-ray bursts, provided that they take place within the
ergosphere. As elaborated in FK08, the resulting QPOs are different
in character to those examined so far in the literature in that they
do not require any modulation/oscillation as such in the signal. The
difference in their character can be easily assessed in the data by
calculating, in addition to the PSDs, also the ACFs, which in the
present case has the form shown in Figures~\ref{fig:obs-dep} and
\ref{fig:src-dep}, i.e. a peak near zero (self-coherence) with a
second one at $\tau \simeq 14M = 0.73 M_1$ msec (QPO), very
different from that of an oscillatory form, corresponding to QPOs
due to some oscillation in the X-ray flux (Fukumura, Dong, Kazanas,
\& Shrader, in preparation). In summary, for sources
concentrated near the equator (e.g. $\theta_s \lsim 85^\circ$), we
find the presence of
persistent QPOs (and a series of harmonics) of $\nu_{QPO} \sim 1.4$
kHz for a $10\Msun$ black hole (equivalently $\sim 1.4$ mHz for a
$10^7 \Msun$ AGN) for a wide range of line-of-sight viewing angles
(from $\sim 30^\circ$ to $\sim 80^\circ$). This QPO
signal is the direct consequence of photon trajectories undergoing
multiple rotations around the black hole due to its strong
frame-dragging effects near the
equator. As the source extends toward the mid-latitudinal region
($\sim 53^\circ$) the QPO signal is weakened by more dominant
incoherent signals.

The QPOs proposed in this work should be viewed as a new class of
QPOs in view of their (i) expected frequency bandpass and (ii) their
underlying mechanism, generic to the dragging of inertial frames on
individual photons. As such, we do not believe these are associated
with the observed HFQPOs often detected in accreting black hole
systems (whose frequency is lower by a factor of $\sim 5$ and some
of which exhibit a 2:3 frequency commensurability that the present
model does not provide for). The fundamental requirement for the
presence of QPOs of the type discussed herein is that the X-ray
flares would take place within the ergosphere of a rapidly-rotating
black hole. The requirement that the ISCO lies within the
ergosphere, then, implies that the black hole spin be $a/M \gtrsim
0.94$. While this appears to be a rather strong constraint, one
should point out that broad Fe line fits by emission from thin
accretion disks around Kerr black holes, provide consistently
estimates of the spin parameter near $a/M \sim 0.99$
\citep[e.g.][]{Brenneman06}. This value is consistent with that used
in the present work.

The above results are relevant if the accreting gas surrounding the
central engine is relatively tenuous and optically transparent to
radiation from these X-ray sources. For example, this should be the
case for optically-thin ADAFs where the accreting plasma
distribution is quasi-spherical (though most of emission is still
originating from the midplane of the gas), in contrast to the
equatorial structure of a standard, thin accretion disk
\citep[e.g.][]{SS73,NT73}. Such geometrically thick, optically thin
configurations appear to be associated with objects such as our
Galactic Center source Sgr A$^\ast$ \citep[e.g.][]{Yuan03,Meyer06},
and Low Ionization Nuclear Emission-line Region sources or LINERs
(e.g. NGC~4258; see Quataert 1999) and nearby giant ellipticals in
low-luminosity AGNs or LLAGNs (e.g. M87; see Di Matteo et al.~2003).
%
%
If the central emission region in these sources is indeed
optically-thin characterized by ADAFs or its more generalized class
of radiatively-inefficient accretion flows (RIAFs; see, e.g. Yuan et
al.~2003), then one may expect a transparent environment
in which photons can escape the production region without much
additional scattering.

QPO features have recently been associated with the PSD of AGN: {\it
XMM-Newton} observations of the the bright nearby Seyfert 1 galaxy
Mrk~766, known to be a NLS1, exhibited statistically significant
peak in the PSD of its light curve \citep[][]{Markowitz07}, modeled
as Lorentzian of $\nu_{QPO} \sim 0.46$ mHz, similar to QPOs detected
from other black hole binary systems. Assuming an estimated mass of
the Mrk~766 nucleus ($\sim 5 \times 10^6 \Msun$), suggested from
optical H$\beta$ emission line velocity dispersion
\citep[e.g.][]{Wandel02}, we can estimate the frame-dragging QPO
frequency to be $\nu_{QPO} \sim 2.7$ mHz, i.e. a factor of $\sim 5$
higher than the observed frequency. It is possible that other NLS1s
with slightly more massive nuclei (order of $10^7\Msun$) may exhibit
a QPO frequency that can be explained in the context of our model.
For example, a radio-loud NLS1, PKS~0558-504, shows a rapid X-ray
flare presumably with a heavier nucleus of $M \sim 4.5 \times
10^7\Msun$ \citep[e.g.][]{Wang01} which may make this source a good
candidate for testing our prediction. Recently, \citet{Meyer06}
discussed a characteristic time-lag due to higher-order emission
from our Galaxy Center (Sgr A$^*$), which may also be a promising
target for our investigation (because of larger mass and longer
timescales compared to stellar-mass black holes). In addition to
Seyfert 1s above, perhaps intermediate mass black holes possibly
associated with ULXs may also be promising sites to search for the
QPOs proposed herein \citep[e.g.][]{Strohmayer07}. Although it may
not be observationally easy to disentangle a potential QPO signal
from high luminosity continuum (presumably at near-Eddington rate)
from these objects, it remains to be studied.


Our discussion so far has been confined to X-ray flare sources that
orbit the black hole in the azimuthal direction with no significant
radial motion. One could argue, however, that any source in the
vicinity of an accreting black hole is likely to be plunging-in
radially at a good fraction of $c$. This is certainly the case for
sources interior to the ISCO. As pointed out above, in order that a
source be both at the ISCO and within the black hole ergosphere, the
spin parameter $a$ should be $a/M \gtrsim 0.94$. For more moderate
values of the spin parameter, say, $a/M \simeq 0.9$, the ISCO is
outside the ergosphere and plasma traversing the latter should have
considerable radial velocity in addition to its azimuthal motion.
Since there is no reason for which this plasma could not produce
X-ray flares (for example, shocked plasma in plunging regions can in
principle serve as a good candidate for such a thermal/nonthermal
X-ray sources, see Fukumura et al.~2007), the entire analysis
described in this work can be extended to this circumstance too. It
is expected, however, that because of its (substantial) radial
velocity, the emitted photons would be Doppler beamed in the radial
direction too and a large number of them would be lost through the
black hole horizon; as a result, the QPO production efficiency may
decrease to an unobservable level for sufficiently small values of
the spin parameter $a$. A detailed study of this arrangement and the
limiting value for which this approach can lead to QPO signals is
deferred to future work.

A related issue is that of the detectability of QPOs produced in the
way described above. In this work we neglected the background
distribution of signal/noise associated with other physical
processes (e.g. emission from accreting flows and/or corona), photon
counting statistics (Poisson noise) and also instrumental responses
to the signal. Depending on the QPO signal strength relative to the
externally contaminating (e.g. accreting gas and coronal) emission
intensity, the induced light echo can be sufficiently weakened to a
statistically indiscernible level. To assess a potential
observability of the predicted QPO signal in this scenario we need
to combine the synthetic signal with the stochastic noise. If
fraction of the observed X-rays does not participate in the
production of the light echoes discussed herein, one should add to
our synthetic signal an underlying constant (stable) component with
some signal-to-noise ratio. Finally, in order to gauge the
observability of the signal we propose, one should also add the
noise associated with photon statistics and detector background.
Such a detail treatment of the observability of the QPOs proposed in
this work is deferred to another future work.

\acknowledgments

We would like to thank the anonymous referee for a number of useful
and insightful suggestions. K.F. is grateful to Fotis Gavriil for
helpful discussions about timing analysis.

\section{Appendix A}

In this Appendix we describe in details the notion of local isotropy
and derive the equations necessary to calculate the subsequent
photon trajectories in Kerr geometry.

In order to correctly describe local isotropy in emission from a
source orbiting at relativistic speed, we first consider the
differential photon distribution as a function of the angle $\psi$
between the source direction and the photon emission in the fluid
frame
\begin{eqnarray}
\frac{dN}{d\psi} = \frac{dN}{d(\cos \psi)} \frac{d(\cos
\psi)}{d\psi} \  . \label{eq:A1}
\end{eqnarray}
The notion of local isotropic emission is the requirement that the
number of photons per local solid angle should be the same for all
the direction of emission, i.e. that
\begin{eqnarray}
\frac{dN} {d\Omega} \equiv N_0 \ , \label{eq:A2}
\end{eqnarray}
where $N_0$ denotes the number of photons within the solid angle
$d\Omega$ which must be conserved from one frame to another and
$d\Omega = 2\pi \sin \psi d\psi$. Therefore, we obtain
\begin{eqnarray}
\frac{dN}{d\psi} = 2\pi N_0 \sin \psi \ , \label{eq:A3}
\end{eqnarray}
which provides the weighting factor as a function of $\psi$.

In the LNRF (or lab frame), where the photon trajectories are
computed, we similarly find
\begin{eqnarray}
\frac{dN}{d\psi'} = \frac{dN}{d(\cos \psi)} \frac{d(\cos
\psi)}{d(\cos \psi')} \frac{d(\cos \psi')}{d\psi'} \ , \label{eq:A4}
\end{eqnarray}
where again
\begin{eqnarray}
\frac{dN}{d(\cos \psi)} = 2 \pi N_0 \ . \label{eq:A5}
\end{eqnarray}
Using equation~(\ref{eq:boost}), we can express
equation~(\ref{eq:A4}) as
\begin{eqnarray}
\frac{dN}{d\psi'} = 2\pi N_0 f(\beta,\psi) \ , \label{eq:A6}
\end{eqnarray}
where
\begin{eqnarray}
f(\beta,\psi) \equiv \frac{(1+\beta \cos \psi) \sin
\psi}{(1-\beta^2)^{1/2}} \ , \label{eq:A7}
\end{eqnarray}
Note that $dN/d\psi' \rightarrow dN/d\psi$ as $\beta \rightarrow 0$
and we can see $\int_0^{\pi} (dN/d\psi') d \psi' = \int_0^{\pi}
(dN/d\psi)d \psi = 4\pi N_0$ regardless of the value of $\beta$ as
expected. Hence, $f(\beta,\psi)$ describes the differential photon
distribution per unit opening angle (i.e. the weighting factor) in
general inertial frames.

One can easily convert the above distribution into the differential
photon distribution over a finite bin $\Delta \psi'$ (in the LNRF)
given by equation~(\ref{eq:boost2}). Note that $\Delta \psi'$ is not
constant and one finds
\begin{eqnarray}
\left(\frac{dN}{d\psi'}\right) \Delta \psi' =
\left(\frac{dN}{d\psi}\right) \Delta \psi = 2\pi N_0 \sin \psi
\Delta \psi \ , \label{eq:A8}
\end{eqnarray}
which states that the number of photons emitted within the
corresponding angular bin in the two frames is indeed the same (i.e.
independent of inertial frame), as one expects. Then, it is
guaranteed that for the entire sky
\begin{eqnarray}
N = \sum_{i=1}^n \left(\frac{dN}{d\psi_i'}\right)\Delta \psi_i' =
\sum_{i=1}^n \left(\frac{dN}{d\psi_i}\right) \Delta \psi_i \cong
2\pi N_0 \int_0^{\pi} \sin \psi d \psi = 4\pi N_0 \ , \label{eq:A9}
\end{eqnarray}
where the integer $i$ represents a discretised grid point in the
polar opening angle and $n$ denotes the total number of bins of the
same angle from $0$ to $\pi$. This prescription of photon
distribution allows one to compute the number of photons in each
local opening angle $\psi'$, consistent with isotropic emission at
the fluid frame. For each such polar angle, we then launch photons
in the azimuthal (about the source velocity) direction $\chi$ (see
Fig.~\ref{fig:angle}) between 0 and $2 \pi$ in equal intervals of
width
\begin{eqnarray}
\Delta \chi \equiv 2\pi \left(\frac{dN}{d\psi'} \Delta \psi'
\right)^{-1} = \frac{1}{N_0 \sin \psi \Delta \psi} \ ,
\label{eq:A12}
\end{eqnarray}
with this prescription, then, guaranteeing the proper number and
photon directions in the LNRF consistent with isotropic emission in
the fluid rest frame.

Besides the Lorentz boost of the angles to the LNRF one must in
addition relate these angles to the photon angular momenta
($\xi,\eta$), as these are the quantities needed in the integration
of the photon geodesics. A simple geometry of this problem relates
the opening angle $\psi'$ and the azimuthal angle $\chi$ to the
corresponding polar and azimuthal angles, $\delta'$ and $\gamma'$,
as described in \S 2.3
\begin{eqnarray}
\cos \psi' = \sin \delta' \sin \gamma' ~~~~{\rm and} ~~~~ \cos
\delta' = \sin \psi' \cos \chi \  , \label{eq:A10}
\end{eqnarray}
which is schematically illustrated in Figure~\ref{fig:angle}.

%

\section{Appendix B}

In Appendix A we discuss how to prescribe local isotropy from an
arbitrary X-ray source position in Kerr geometry, which enables us
to calculate the emission direction ($\delta',\gamma'$) of an
individual photon. In order to subsequently integrate the geodesic
equations (photon trajectories) per photon given by
equations~(\ref{eq:t}) through (\ref{eq:phi}) one needs to translate
such a directional information into the photon's angular momentum
(equivalent to impact parameters) in the LNRF. Since we are dealing
with 3D null geodesics each photon is characterized by two constants
$\xi$ and $\eta$ where $\xi$ is the axial component of angular
momentum and $\eta$ is closely related to the polar component of
angular momentum $p_\theta$ by $\eta \equiv p_\theta^2+\xi^2 \cot^2
\theta -a^2 \cos^2 \theta$ \citep[e.g.][]{Bardeen72,Chandra83}. Here
we derive analytic expressions in the most general case for
($\xi,\eta$) in terms of the local emission angles
($\delta',\gamma'$).

Considering equations~(\ref{eq:eqn1}) and (\ref{eq:eqn2}) where the
left-hand-side is related to the local emission angles and the
right-hand-side to geometry and velocity, one can eliminate the
derivatives ($\dot{t},\dot{r},\dot{\theta},\dot{\phi}$) using
equations~(\ref{eq:t}) through (\ref{eq:phi}) to express the
right-hand-side in terms of ($\xi,~\eta$). By solving these
equations one can obtain analytic forms of
$\xi=\xi(\delta',\gamma')$ and $\eta=\eta(\delta',~\gamma')$
explicitly as
\begin{eqnarray}
\xi(\delta',\gamma') &=& \left[\frac{(1-\mu^2) \omega A \zeta \sin^2
\delta' \pm \left\{(1-\mu^2) \Delta \Sigma^2 \zeta^2 \sin^2 \delta'
\csc^2 \gamma' \right\}^{1/2}}{\Delta \Sigma^2 \csc^2
\gamma'-(1-\mu^2) \omega^2 A^2 \sin^2 \delta'}\right]_{\rm source} \
, \label{eq:B1}
\\
\eta(\xi;\delta',\gamma') &=& \left[\frac{\mu^2 \zeta \Delta
\{\xi^2-a^2(1-\mu^2)\} +(1-\mu^2) (\zeta-\omega A \xi)^2 \cos^2
\delta'}{(1-\mu^2) \zeta \Delta}\right]_{\rm source} \  ,
\label{eq:B2}
\end{eqnarray}
where $\zeta \equiv r^4+a^2 \{\mu^2 \Delta+ r(r+2M)\}$ and $\mu
\equiv \cos \theta$ with all the quantities to be evaluated at the
source position $(r_s,\theta_s,\phi_s)$. The plus/minus sign for the
second-term in the numerator of equation~(\ref{eq:B1}) depends on
the initial direction of photons emitted in such a way that $\xi
\geq 0$ for all emission; i.e., plus for photons initially emitted
(locally) in the forward direction ($0 \leq \gamma' \le \pi$) and
minus for backward direction ($\pi \leq \gamma' \le 2\pi$) with
respect to the LNRF (which can be in rotation relative to the
observer). Note that there is an ``offset" factor in directional
information due to this rotation of local inertial frame [due to the
quantity $\omega$ in eqns.~(\ref{eq:B1}) and (\ref{eq:B2})]. In the
absence of frame-dragging ($a/M \rightarrow 0$) we see that
\begin{eqnarray}
\xi \rightarrow \pm r \left(\frac{1-\mu^2}{1-2M/r}\right)^{1/2} \sin
\delta' \sin \gamma' \ , \label{eq:B3}
\end{eqnarray}
which is positive for both forward and backward emission with $\sin
\delta' \ge 0$, as expected (see Fig.~\ref{fig:angle}). This also
reproduces correct axial component $\xi$ in Newtonian geometry ($r
\rightarrow \infty$); when emission is exactly radial, either away
($\gamma'=0$) or toward ($\gamma'=\pi$) the rotation (symmetry)
axis, $\xi$ will vanish as required. When emission is completely
parallel to the rotational axis (either $\delta'=0$ or $\pi$), one
also recovers $\xi \rightarrow 0$.

As a special case we can see that for emission from an on-axis
(z-axis) source, i.e. as $\theta \rightarrow 0 ~ (\mu \rightarrow
1)$
\begin{eqnarray}
\xi \rightarrow 0 ~~ {\rm and} ~~ \eta \rightarrow {\rm finite} \ ,
\label{eq:B4}
\end{eqnarray}
indicating the absence of axial angular momentum, as anticipated.
For another special case where emission is confined in the
equatorial plane or (x,y)-plane as $\theta \rightarrow \pi/2 ~ (\mu
\rightarrow 0)$ and $\delta' \rightarrow \pi/2$ we see
\begin{eqnarray}
\xi \rightarrow {\rm finite} ~~{\rm and} ~~ \eta \rightarrow 0 \ ,
\label{eq:B5}
\end{eqnarray}
where $\eta$, which in this case is exactly $p_\theta^2$, vanishes,
also as expected.

\end{document}